\documentclass{aa}

\usepackage{graphicx}
\usepackage{natbib}
\bibpunct{(}{)}{;}{a}{}{,} 
\DeclareGraphicsExtensions{.eps,.ps}
\usepackage{txfonts}
\usepackage{amssymb}
\usepackage{xspace}
\usepackage{ifpdf}

\ifpdf
\usepackage{epstopdf}
\usepackage[colorlinks=true,linkcolor=magenta,citecolor=dgreen,breaklinks=true]{hyperref}
\fi

\def\eso{ESO\,359-G19\xspace}
\def\he{HE\,1143-1810\xspace}
\def\cts{CTS\,A08-12\xspace}
\def\mkn{Mrk\,110\xspace}

\def\arcminpoint{$'\!.$}

\def\pn{EPIC-pn\xspace} 
\def\mos{EPIC-MOS\xspace}
\def\rgs{RGS\xspace}

\def\kms{\,\textnormal{km~s}\ensuremath{^{-1}}\xspace}
\def\kev{\,\textnormal{keV}\xspace}

\def\hubble{\,\textrm{km~s}\ensuremath{^{-1}}~\textrm{Mpc}\ensuremath{^{-1}}\xspace}

\newcommand{\chired}{\mbox{$\chi^{2}_{\nu}$}}
\def\funits{\textrm{erg~cm}\ensuremath{^{-2}}~\textrm{s}\ensuremath{^{-1}}\xspace}
\def\funitsa{\textrm{erg~cm}\ensuremath{^{-2}}~\textrm{s}\ensuremath{^{-1}}~\textrm{\AA}\ensuremath{^{-1}}\xspace}
\def\funitsk{\textrm{erg~cm}\ensuremath{^{-2}}~\textrm{s}\ensuremath{^{-1}}~\textrm{keV}\ensuremath{^{-1}}\xspace}

\def\normunits{\textrm{ph}~\textrm{cm}\ensuremath{^{-2}}\,\textrm{s}\ensuremath{^{-1}}\,\textrm{keV}\ensuremath{^{-1}}\xspace}

\usepackage{color}

\definecolor{dgreen}{rgb}{0,.5,.1} 
\definecolor{pink}{rgb}{.9,.2,.5}  
\definecolor{orange}{rgb}{.9,.4,0} 
\definecolor{darkred}{rgb}{.545,0.0,.0}
\graphicspath{{figures/}}
\begin{document}

\titlerunning{Comprehensive analysis of the
  \textit{XMM-Newton} data of Sy\,1 galaxies}
\authorrunning{M.V. Cardaci et al.}
\title{A comprehensive approach to analyzing the \textit{XMM-Newton}
  data of Seyfert 1 galaxies}

\author{M\'onica V. Cardaci \inst{1,2,3}, Mar\'ia Santos-Lle\'o \inst{2},  Guillermo F. H\"agele \inst{1,3}\thanks{CONICET, Argentina},
        Yair Krongold \inst{4},  
        \'Angeles I. D\'iaz \inst{1}, Pedro Rodr\'iguez-Pascual \inst{2}} 

\offprints{monica.cardaci@uam.es}

\institute{1 Universidad Aut\'onoma de Madrid, Ctra.\ de Colmenar Km.15,
              Cantoblanco, 28049 Madrid, Spain \\ 
           2 XMM-Newton Science Operations Center, ESAC, ESA, POB 78, E-28691
              Villanueva de la Ca\~nada, Madrid, Spain \\
	   3 Facultad de Cs.\ Astron\'omicas y Geof\'isicas, Universidad
              Nacional de La Plata, Paseo del Bosque s/n, 1900 La Plata, 
              Argentina \\
           4 Instituto de Astronom\'ia, Universidad Nacional Aut\'onoma de
              M\'exico, Apartado Postal 70-264, 04510 M\'exico DF, M\'exico
           }

\date{ }

\abstract
{}
{We seek a comprehensive analysis of all the information provided by the 
  {{\em XMM-Newton}} 
  satellite of the four Seyfert 1 galaxies \eso, \he,
  \cts, and \mkn, including the UV range, to characterize the different
  components that are emitting and absorbing radiation in the vicinity of the
  active nucleus.}
{The continuum emission was studied through the EPIC spectra by taking
  advantage of the spectral range of these cameras. The high-resolution RGS
  spectra were analyzed to characterize the absorbing and
  emission line features that arise in the spectra of the sources. 
  All these data, complemented by information in the UV, are analyzed jointly
  in order to achieve a consistent characterization of the observed features
  in each object.}
{The continuum emission of the sources can be
  characterized either by a combination of a power law and a black body for the
  weakest objects 
  or by two power law components for the brightest ones. 
  The continuum is not absorbed by neutral or ionized material in the line of
  sight to any of these sources. In all of 
  them we have identified a narrow Fe-K$\alpha$ line at 6.4\,keV. 
  In \eso we also find an Fe{\sc xxvi} line at about 7\,keV. In the soft
  X-rays band, we identify only 
  one O{\sc vii} line in the spectra of \he and \cts, and two O{\sc
  vii}-He$\alpha$ triplets and a narrow O{\sc viii}-Ly$\alpha$
  emission line in \mkn.
} 
{
  Not detecting warm material in the line of sight to the low state
  objects is due to intrinsically weaker or absent absorption in the line of
  sight and not to a low signal-to-noise ratio in the data. 
  Besides this, the absence of clear emission lines cannot be fully
  attributed to dilution of those lines by a strong continuum. 
}


\maketitle

\section{Introduction \label{sec:intro}}

\medskip

The spectral features characteristic of the X-ray spectra of active galactic
nuclei (AGNs) are conspicous signatures of the physical processes taking place
in the inner regions of these objects, and analyzing them has significantly
improved our understanding of these issues in the last few years of the 
20th century. However, it has also become evident that a much better
understanding is needed, and in fact it can be achieved by detailed analysis
of the data acquired with the present generation of X-ray telescopes.

In this paper we present a comprehensive analysis of all the available data
obtained with the {\em XMM-Newton} satellite of four Seyfert
1 galaxies: \eso, \he, \cts, and \mkn in order to try to characterize the
different components of the gas that imprint the absorption and emission 
features in the X-ray spectra of these objects. 
The low-resolution and large spectral range EPIC (European Photon Image
Camera) spectra were used to study the continuum emission. 
The high-resolution RGS (Reflection Grating Spectrometer) spectra are
analyzed to search for and characterize, if found, the absorbing
and emission line features that arise in the spectra of these
sources. 

\begin{table*}[ht]
{\footnotesize
\caption{General characteristics of the objects selected for our study.}
\label{objects}
\centering
\begin{tabular}{@{}l cc c c c@{}}
\hline\hline
\noalign{\smallskip}			  
Object & $\alpha_{2000}$ & $\delta_{2000}$ & z               & Distance (Mpc)$^*$
& \textit{N}$_{\mathrm H}$ (cm$^{-2}$) \\
\noalign{\smallskip}			  
\hline
\noalign{\smallskip}			  
\eso & 04$^h$ 05$^m$ 01.7$^s$ & -37$^\circ$ 11' 15\arcsec0 & 0.056$^{(a)}$ & 224 & 1.02$\times 10^{20}$ \\
\he  & 11$^h$ 45$^m$ 40.5$^s$ & -18$^\circ$ 27' 16\arcsec0 & 0.033$^{(b)}$ & 132 & 3.40$\times 10^{20}$ \\
\cts & 21$^h$ 32$^m$ 02.2$^s$ & -33$^\circ$ 42' 54\arcsec0 & 0.029$^{(c)}$ & 116 & 4.07$\times 10^{20}$ \\
\mkn & 09$^h$ 25$^m$ 12.9$^s$ & +52$^\circ$ 17' 11\arcsec0 & 0.035$^{(d)}$ & 140 & 1.42$\times 10^{20}$ \\
  \hline
\noalign{\smallskip}			  
\end{tabular}

$^*$From the object redshifts using the expression 
$D = (cz/H_0)$ and adopting $H_0=75\hubble$. Estimated hydrogen column
densities from \cite{1990ARA&A..28..215Dickey}.
References: 
$^{(a)}$\cite{1996A&AS..115..235Reimers},
$^{(b)}$\cite{1998AJ....116....1DaCosta}, 
$^{(c)}$\cite{2000ApJS..126...63Rodriguez-Ardila}, 
$^{(d)}$\cite{1992ApJS...80..109Boroson}, 
}
\end{table*}

The first of our objects, \eso, has been classified as a Seyfert 1 
galaxy by \cite{1992RMxAA..24..147Maza} using longslit spectrophotometric
data. It has not been studied in detail in spite of being observed on 
many occasions
\citep{1988MNRAS.233..691Fairall,1996A&AS..115..227Wisotzki,1996AJ....111..794Karachentsev}. 
It was previously observed in X-rays with the {\em ROSAT} \citep[1990
 August;][]{1999A&A...349..389Voges,1998A&A...335..467Thomas} and {\em
 Einstein} 
\citep{1992ApJS...80..257Elvis} satellites, as part of the study of the general
properties of objects with X-ray emission
\citep{1997ApJ...489..615Simcoe,2001A&A...367..470Grupe,2003MNRAS.346..304Jahnke,2003A&A...403..247Fuhrmeister}.
More recently, it has been observed by {\em SWIFT} (2008 October) as part of a
sample of 92 
bright soft X-ray selected AGNs \citep{2010ApJS..187...64Grupe}.

\he was classified as a type 1 Seyfert AGN by
\citet{1986ApJ...301..742Remillard}. 
These authors also found that this galaxy has a companion object,
LEDA\,867889, which shows neither signatures of nuclear activity 
nor the characteristic emission of H{\sc ii} galaxies.
It is located at about 20 kpc from \he. 
From extreme UV observations with {\em FUSE}, \cite{2003ApJS..146....1Wakker}
studied the  gas surrounding the AGN and concluded that the
positive-velocity O{\sc vi} absorption extends much further than expected for a
Galactic origin, although they did not find a clear separation between an
intrinsic O{\sc vi}
line and the Galactic absorption. The authors separated
the thick disk and high-velocity cloud components at a velocity of
100\,\kms. However, using the same and newer observations from {\em FUSE},
\cite{2007AJ....134.1061Dunn} could not 
identify any intrinsic absorption feature in this galaxy. 

Very little is known in detail about \cts, despite its
being included in a large number of statistical studies at different
wavelengths
\citep[e.g.][]{1999A&A...349..389Voges,2000ApJS..129..547Bauer,2003A&A...411L..59Ebisawa,2007A&A...472..705Verrecchia,2007MNRAS.375..931Mauch}.
It was classified as a Seyfert 1 by \cite{1994RMxAA..28..187Maza}. 

Finally, \mkn is a nearby Seyfert 1 galaxy with a very irregular morphology
\citep{1999A&A...345...49Bischoff}, which may indicate that it has suffered
some kind of interaction with a companion 
galaxy \citep{1988AJ.....95..677Hutchings}. 
From optical spectroscopy covering a narrow spectral range, between 4300 and
5700\,\AA, \cite{1992ApJS...80..109Boroson} estimated a full width at half
maximum (FWHM)  of 2120\,\kms
for the H$\beta$ emission line, and observed a very broad He{\sc i} 4471\AA\
emission line. More recently, \cite{2010ApJS..187...64Grupe} have derived 
a FWHM of 1760\,$\pm$\,50 \kms for H$\beta$, less than the value used as
one of the criteria to distinguish Narrow Line Seyfert 1 (NLS1) from Seyfert 1
galaxies. Nevertheless, the objects was classified as a NLS1 by 
\cite{2006A&A...455..773Veron}. As in the case of UGC\,11763 
\citep{2009A&A...505..541Cardaci}, \mkn seems to be located in a transition
zone between the Seyfert 1 and NLS1 classifications, showing some of the
properties of each class. 

In Table \ref{objects} we list the general characteristics of the galaxies in
our sample as obtained from the literature: 
celestial coordinates ($\alpha_{2000}$ and $\delta_{2000}$),
redshifts (z), the distances in Mpc as derived from their redshifts using the
expression $D = (cz/H_0)$ \nocite{1997iagn.book.....Peterson} and adopting 
$H_0=75\hubble$,
and estimated Galactic hydrogen column densities (\textit{N}$_{\mathrm H}$)
in the line of sights taken
from \cite{1990ARA&A..28..215Dickey}. 

The paper is organized as follows. In Section \ref{sec:obs} we describe the
observations and data reduction. The optical-UV results and the X-ray spectral
analysis are presented in Sections \ref{sec:opt-uv} and \ref{sec:x-rays},
respectively. In Section \ref{sec:variability} we analyze the variability
during the {\em XMM-Newton} observation and compare our observed fluxes with
those from the literature. Our results are discussed in Section
\ref{sec:discussion}. And finally, the summary and conclusions of this work
are given in Section \ref{conclusions}. 


\section{Observations \label{sec:obs}}

\eso, \he, \cts, and \mkn were observed by {\em XMM-Newton}
\citep{2001A&A...365L...1Jansen} in 2004 
(Table \ref{4obj:journal}) as part of
a project to characterize the complex profile of the iron K$\alpha$ line and
the features that sometimes appear in the soft X-ray spectral range. 
The four galaxies of this study were selected to be brighter
than 0.85 c\,s$^{-1}$ in the {\em ROSAT} Bright Source Catalog
\citep{1999A&A...349..389Voges} and not previously observed by the
\textit{XMM-Newton}. Exposure times were estimated from the available 
{\em ASCA} and {\em ROSAT} observations. 
The observing modes for 
\pn and \mos1 \citep{2001A&A...365L..27Turner} were selected according
to the expected count rates 
(from PIMMS\footnote{http://cxc.harvard.edu/toolkit/pimms.jsp}) 
to avoid pile-up effects, and the selections are listed in Table
\ref{4obj:tabobs}.  
\mos2 camera was always set in full frame mode to allow cross check for
bright serendipitous X-ray sources in the field that could contaminate the
RGS spectra.
RGSs \citep{2001A&A...365L...7DenHerder} were run in the default
spectroscopy mode.
Optical Monitor \citep[OM,][]{2001A&A...365L..36Mason} observations that
combined broad-band imaging filters were requested 
to investigate the circumnuclear structure in the UV domain with a series of
UV-Grism exposures to obtain UV spectral and variability information about the
active nucleus. 

The OM, when used with broad-band filters, was always operated with the
`science user defined' windows mode. The OM windows were centered on the
target, with no spatial binning to maximize the resolution and with the
maximum allowed size for those windows that is 
5\arcminpoint1$\times$5\arcminpoint0. This mode is the best choice for
galaxies that are small enough to be fully included in a few arcmin square
window. The default OM windows are designed to provide data over the whole OM
field of view which, because it is larger, needs more time to be fully
covered. The default grism window was selected for using the OM with the UV
grism. This window contains the zeroth-order image and the first-order
spectrum of the target at full detector resolution. 
More details on the EPIC and OM operating modes can be found in the
\cite{XMM-Users-Handbook2010}.

\begin{table}[ht]
{
\caption{Journal of observations.}
\label{4obj:journal}
\centering
\begin{tabular}{@{}l  c c@{} }
  \hline\hline
  \noalign{\smallskip}			  
  \multicolumn{1}{c}{Object}
   & \multicolumn{1}{c}{Observ. ID}
   & \multicolumn{1}{c}{{Date}} \\[-2pt]
  \multicolumn{1}{c}{}
   & \multicolumn{1}{c}{}
   & \multicolumn{1}{c}{}  \\[2pt]
  \hline
  \noalign{\smallskip}			  
  \eso\ & 02011301010 & 09 March 2004 \\
  \he\  & 02011302010 & 08 June   2004 \\
  \cts\ & 02011303010 & 30 October   2004 \\
  \mkn\ & 02011305010 & 15 November   2004 \\
  \hline
\end{tabular} \\
}
\end{table}

The data were processed with the 
8.0.0 version of the Science Analysis Subsystem 
(SAS) software package \citep{2001A&A...365L...1Jansen} 
using the calibration files available in January 2009. 
All the standard procedures and screening criteria were followed
for extracting the scientific products. 


\subsection{Selection of low background intervals}
High background time intervals have been excluded
using the method that maximizes the signal-to-noise in the spectrum (MaxSNR) 
as described in \citet{2004MNRAS.351..161Piconcelli}. 
The observations of \eso and \cts were affected by an increase in
the background radiation towards the end of the observation. 
The periods affected by high radiation were not taken into account for
the \mos spectral extraction. There was no need, however, to reject any
time interval from the \pn data of any of either source.
For \eso, time intervals with background
count rates higher than 0.6 and 3.2\,c\,s$^{-1}$ had to be discarded from
the \mos1 and \mos2 spectra, respectively.
The total discarded time was about 5\% for \mos1 and about 1.5\% for \mos2
of the total observing time.
In the case of \cts, time intervals with background count rates higher than
0.9 and 0.5 \,c\,s$^{-1}$ were subtracted of \mos1 and \mos2, respectively. 
The total discarded time was about 2.5\% and much less than the 1\% of the total
observing time for \mos1 and \mos2, respectively.

No high background intervals occurred during the \he and \mkn observations. 
Therefore, the \pn spectrum of \he and all the EPIC spectra of \mkn
did not need to be cut in time. Nevertheless, as a result of applying
the MaxSNR procedure, time intervals with
background count rates higher than 0.4\,c\,s$^{-1}$ had to be discarded from
both \mos spectra of \he. The total discarded time was about 1.5\% of
the total observing time in each case.

The light curves of the RGS background radiation corresponding to our data
show a significant increase towards the end of the \eso observation,
reaching even $>1$ c\,s$^{-1}$ for RGS2, whose sensitivity is slightly
greater than that of RGS1. 
Aproximately the first 19\,ks of the observation had a background
count rate below 0.2 c\,s$^{-1}$, and then it increased (in average) to
about 0.4 c\,s$^{-1}$. During the observations of \he and \mkn, the
background count rates of the RGS spectra were less than 0.2 c\,s$^{-1}$. 
Despite the increase in the background radiation that occurred during the
\eso 
and \cts observations, it was unnecessary to exclude the 
periods of increased background radiation since it always remained
between 0.1 and 1.3 counts per second, which is considered normal 
\citep{XMM-SAS-Users-Guide}.

Detailed instruments configurations are given in Table \ref{4obj:tabobs} 
for each object. 
The last column of this table lists the final effective exposure
time after taking the live time\footnote{The live time is 
the ratio between the time interval during which the CCD is collecting X-ray
events (integration time, including any time needed to shift events towards the
readout) and the frame time (which in addition includes time needed for the
readout of the events). {\em XMM-Newton} Users Handbook 
({\tt http://xmm.esac.esa.int/external/xmm\_user\_support/
documentation/uhb/node28.html}).} into account. 

\begin{table}[htbp]
\begin{minipage}[t]{\columnwidth}
\caption{Details of {\em XMM-Newton} instrument exposures.}
\label{4obj:tabobs}
\centering
{\tiny
\begin{tabular}{@{}lllcc@{}}
\hline\hline
\noalign{\smallskip}
Instrument & Mode & Filter & Time (s) & Time (s) \\
           &      &        & Scheduled & Effective \\
\hline 
\noalign{\smallskip}
\multicolumn{5}{c}{\textbf{\eso}} \\
EPIC-pn   & Small Window & Thick        & 23964 & 16804$^{\mathrm{a}}$ \\
EPIC-MOS1 & Small Window & Thin         & 24157 & 19375 \\
EPIC-MOS2 & Full Frame   & Medium       & 24172 & 23908 \\
RGS1      & Spectroscopy & --           & 24385 & 23968 \\
RGS2      & Spectroscopy & --           & 24380 & 23953 \\
OM        & Image$^{\mathrm{b}}$& U            &  1300 & \\
OM        & Image$^{\mathrm{b}}$& U            &  1200 & \\
OM        & Image$^{\mathrm{b}}$& B            &2$\times$1200 & \\
OM        & Image$^{\mathrm{b}}$& UVW1         &2$\times$1400 & \\
OM        & Image$^{\mathrm{b}}$& UVM2         & 1500 & \\ 
OM        & Image$^{\mathrm{b}}$& UVM2         & 1400 & \\ 
OM        & Image$^{\mathrm{b}}$& UVW2         &2$\times$1800 & \\
OM        & Image$^{\mathrm{b}}$& UV Grism     &4$\times$1400 & \\
\noalign{\smallskip}
\hline 
\noalign{\smallskip}
\multicolumn{5}{c}{\textbf{\he}}\\
EPIC-pn   & Small Window & Thin         & 30964 & 21717$^{\mathrm{a}}$ \\
EPIC-MOS1 & Small Window & Thin         & 31157 & 29660 \\
EPIC-MOS2 & Full Frame   & Medium       & 31172 & 30253 \\
RGS1      & Spectroscopy & --           & 31385 & 30913 \\
RGS2      & Spectroscopy & --           & 31380 & 30887 \\
OM        & Image$^{\mathrm{b}}$& U            &2$\times$1200 & \\
OM        & Image$^{\mathrm{b}}$& B            &2$\times$1300 & \\
OM        & Image$^{\mathrm{b}}$& UVW1         &2$\times$1300 & \\
OM        & Image$^{\mathrm{b}}$& UVM2         &2$\times$1400 & \\ 
OM        & Image$^{\mathrm{b}}$& UVW2         &2$\times$1500 & \\
OM        & Image$^{\mathrm{b}}$& UV Grism     &2$\times$1200 & \\
OM        & Image$^{\mathrm{b}}$& UV Grism     &7$\times$1300 & \\
\noalign{\smallskip}
\hline 
\noalign{\smallskip}
\multicolumn{5}{c}{\textbf{\cts}}\\
EPIC-pn   & Small Window & Thin         & 45964 & 32089$^{\mathrm{a}}$ \\
EPIC-MOS1 & Small Window & Thin         & 46157 & 42584 \\
EPIC-MOS2 & Full Frame   & Medium       & 46172 & 44787 \\
RGS1      & Spectroscopy & --           & 46385 & 45841 \\
RGS2      & Spectroscopy & --           & 46380 & 45817 \\
OM        & Image$^{\mathrm{b}}$& V            &2$\times$1200 & \\
OM        & Image$^{\mathrm{b}}$& U            &2$\times$1200 & \\
OM        & Image$^{\mathrm{b}}$& B            &2$\times$1200 & \\
OM        & Image$^{\mathrm{b}}$& UVW1         &3$\times$1400 & \\
OM        & Image$^{\mathrm{b}}$& UVW1         &2$\times$3200 & \\
OM        & Image$^{\mathrm{b}}$& UVW1         &4$\times$1200 & \\
OM        & Image$^{\mathrm{b}}$& UVM2         &2$\times$2000 & \\ 
OM        & Image$^{\mathrm{b}}$& UVW2         &2$\times$2500 & \\
OM        & Image$^{\mathrm{b}}$& UV Grism     &2$\times$4000 & \\
\noalign{\smallskip}
\hline 
\noalign{\smallskip}
\multicolumn{5}{c}{\textbf{\mkn}$\!^{\mathrm{c}}$}\\
EPIC-pn   & Small Window & Thin         & 46964 & 32861$^{\mathrm{a}}$ \\
EPIC-MOS1 & Small Window & Thin         & 47157 & 45656 \\
EPIC-MOS2 & Full Frame   & Medium       & 47172 & 46554 \\
RGS1      & Spectroscopy & --           & 47385 & 47189 \\
RGS2      & Spectroscopy & --           & 47380 & 47152 \\
OM        & Image$^{\mathrm{b}}$& V            &2$\times$1200 & \\
OM        & Image$^{\mathrm{b}}$& U            &2$\times$1300 & \\
OM        & Image$^{\mathrm{b}}$& B            &2$\times$1300 & \\
OM        & Image$^{\mathrm{b}}$& UVW1         &10$\times$1400 & \\
OM        & Image$^{\mathrm{b}}$& UVM2         &2$\times$2000 & \\ 
OM        & Image$^{\mathrm{b}}$& UVW2         &2$\times$2500 & \\
OM        & Image$^{\mathrm{b}}$& UV Grism     & 4000 & \\
OM        & Image$^{\mathrm{b}}$& UV Grism     & 2160 & \\
\noalign{\smallskip}
\hline 
\noalign{\smallskip}
\end{tabular}}
\end{minipage}
$^{\mathrm{a}}$The live time of the EPIC-pn small window mode is
0.71. $^{\mathrm{b}}$ In the science user defined mode. $^{\mathrm{c}}$
Observation was offset for calibration purposes.
\end{table}


\section{Optical-UV analysis \label{sec:opt-uv}}

OM data were processed with the SAS task {\sc
omichain} and default parameters. In the broad-band OM images, there is no
clear evidence of extended emission.
The images provide flux measures in the selected filters: B, U,
UVW1, UVM2, and UVW2 for all objects and also V for \cts and \mkn.
Table\,\ref{4obj:om-fluxes} shows the effective wavelengths of these filters,
together with measured fluxes.
When more than one exposure per filter is available, the mean value is
listed. The flux differences on each series of consecutive exposures with the
same filter are all compatible with no variability within the measurement
errors ($<$3\%). 

The spectrum recorded in each individual UV-grism exposure is very weak; in
addition, a number of zero-order images near the spectrum location and along
the dispersion further complicate the extraction of the source spectrum. As a
result, the total signal-to-noise of the extracted spectra was not as high as
expected. 

\begin{table*}[ht]
\begin{minipage}[t]{\textwidth}
\caption{Fluxes in the OM filters obtained using aperture photometry.}
 \label{4obj:om-fluxes}
 \centering
 \begin{tabular}{@{}lrrrrrr@{}}
 \hline\hline
 \noalign{\smallskip}
\multicolumn{1}{@{}r}{filter}
   & \multicolumn{1}{c}{V}  &  \multicolumn{1}{c}{B}  &  \multicolumn{1}{c}{U} 
   &  \multicolumn{1}{c}{UVW1}  &  \multicolumn{1}{c}{UVM2}  &  \multicolumn{1}{c}{UVW2} \\
\multicolumn{1}{@{}r}{$\lambda_{\textrm{eff}}$}
   &\multicolumn{1}{c}{5430\,\AA} & \multicolumn{1}{c}{4500\,\AA} & \multicolumn{1}{c}{3440\,\AA} & \multicolumn{1}{c}{2910\,\AA} & \multicolumn{1}{c}{2310\,\AA} & \multicolumn{1}{c}{2120\,\AA} \\
 \noalign{\smallskip}
 \hline
 \noalign{\smallskip}
  \eso &\multicolumn{1}{c}{--}& $1.43\pm0.01$ & $1.34\pm0.01$ & $1.96\pm0.03$ & $1.86\pm0.06$ & $1.7\pm0.10$ \\
  \he  &\multicolumn{1}{c}{--}& $7.65\pm0.02$ & $15.82\pm0.03$ & $19.68\pm0.05$ & $24.4\pm0.10$ & $26.8\pm0.40$ \\
  \cts &$1.78\pm0.02$& $1.60\pm0.02$ & $2.31\pm0.02$ & $2.80\pm0.01$ & $3.00\pm0.07$ & $2.6\pm0.10$\\
  \mkn & $5.75\pm0.03$ & $7.30\pm0.02$ & $14.16\pm0.03$ & $17.53\pm0.02$ & $22.4\pm0.10$ & $24.8\pm0.30$ \\
  \noalign{\smallskip}
  \hline
  \noalign{\smallskip}
\end{tabular}
\end{minipage}\par
Fluxes are the mean values of the multiple exposures on each
filter  and are given in $10^{-15}$\,erg\,\,cm$^{-2}$\,s$^{-1}$\,\AA$^{-1}$
units. 
\end{table*}

\begin{figure}
\centering
\resizebox{\hsize}{!}{\includegraphics{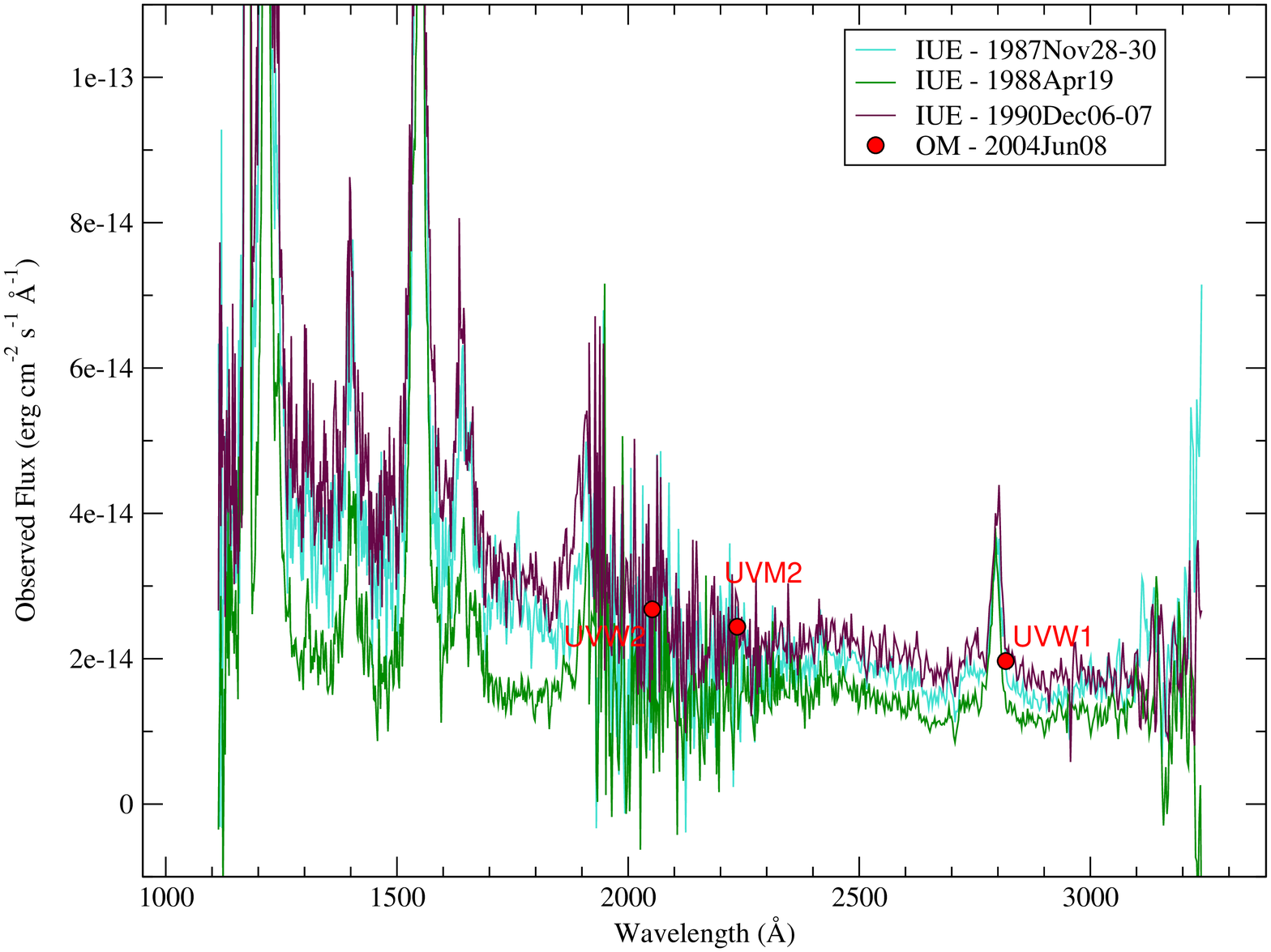}}\\
\vspace*{0.3cm}
\resizebox{\hsize}{!}{\includegraphics{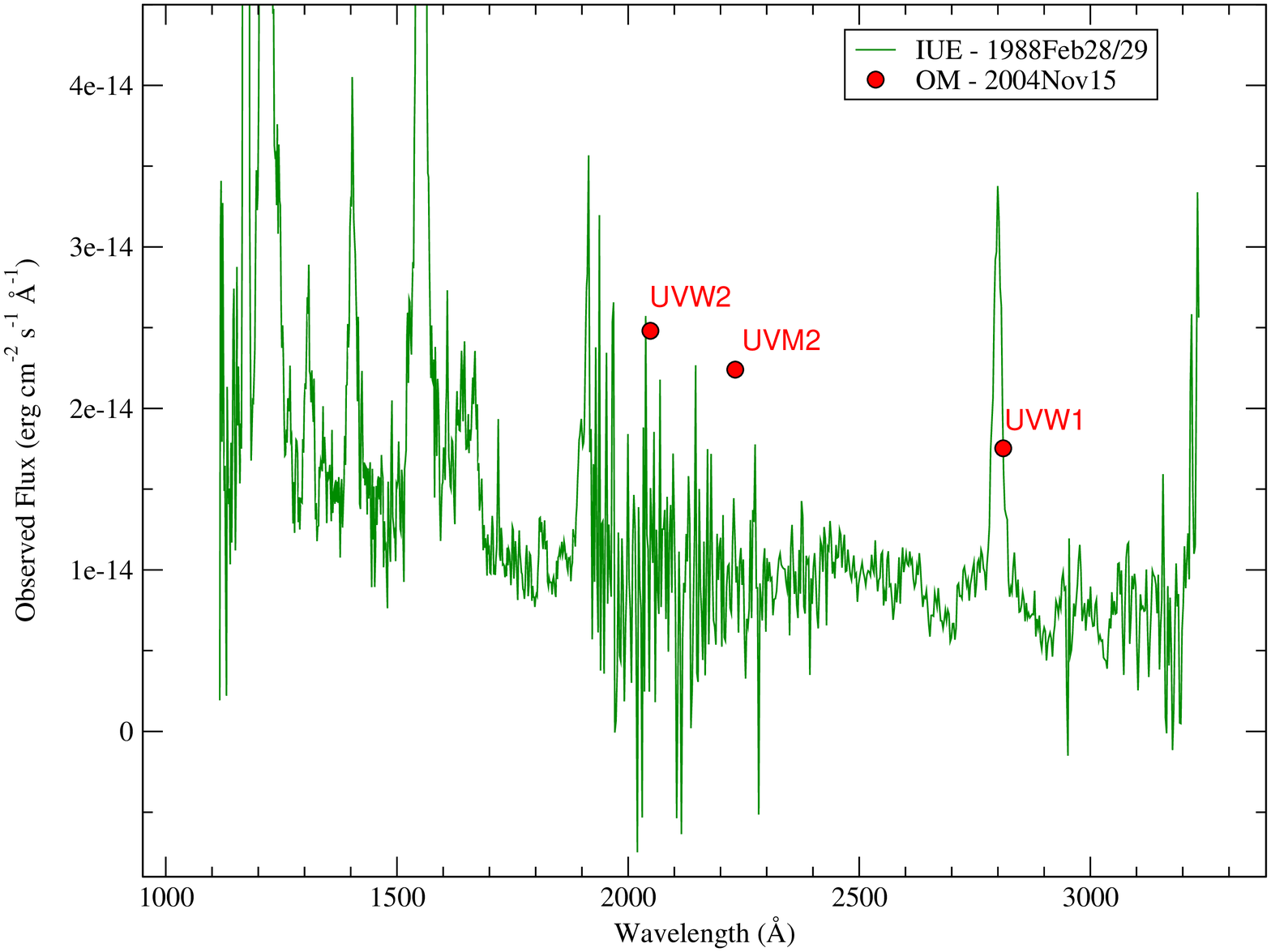}}
 \caption{Average IUE spectra (1200-3200\,\AA) of \he (top) for the three
 observing epochs, and \mkn (bottom) for the observing epoch. 
 Solid red circles show the OM measures for the UVW2, UVM2, and UVW1 filters.} 
 \label{he-mkn-figuvspec}
\end{figure}

\he was observed by the {\em International Ultraviolet Explorer} ({\em IUE}) 
in three different epochs between 1987 and 1990 with its Short Wavelength
(1150--1950\AA) and Long Wavelength (1950--3200\AA)
spectrographs. OM flux measurements in the UVW2, UVM2, and UVW1 filters are 
overplotted in the top panel of Figure \ref{he-mkn-figuvspec} on the average
IUE spectra of each epoch.
In this figure we see that the average UV spectrum taken in 1990 is indeed an
acceptable representation of the UV spectral energy distribution (SED)
of \he at the time of the {\em XMM-Newton} observation.
From the IUE average spectrum in the rest frame we took the UV flux at
2500\,\AA, $F(2500\,\mathrm{\AA})=2.2\times 10^{-14}$\,\funitsa. 
This value has been corrected for neither the Balmer
continuum nor the Fe{\sc ii} contributions.

\mkn was also observed by the {\em IUE} in 1988 with its Short Wavelength
and Long Wavelength spectrographs. Figure \ref{he-mkn-figuvspec} (bottom
panel) shows the average {\em IUE} spectrum with the fluxes obtained from the
OM filters overplotted. This figure shows that \mkn was in a lower flux state
when observed by {\em IUE} than when observed by {\em XMM-Newton}. After
comparing the fluxes obtained for the OM filters UVM2 and UVW1 with those from
the {\em IUE} observation for the same wavelengths, we estimated that the
difference between them is about $1\times10^{-14}$\,\funitsa.
Adding this value to the flux at 2500\,\AA\ obtained from the
\textit{IUE} average spectrum in the rest frame and assuming that the shape of
the UV continuum is essentially the same for the different epochs (as in the
case of \he), we estimate that 
$F(2500\,\mathrm{\AA})=2.05\times10^{-14}$\,\funitsa at the time
of our observation.


\section{Variability \label{sec:variability}}

\begin{figure}
\centering
\resizebox{\hsize}{!}{\includegraphics{16198fg2t.eps}}\\
\vspace*{0.3cm}
\resizebox{\hsize}{!}{\includegraphics{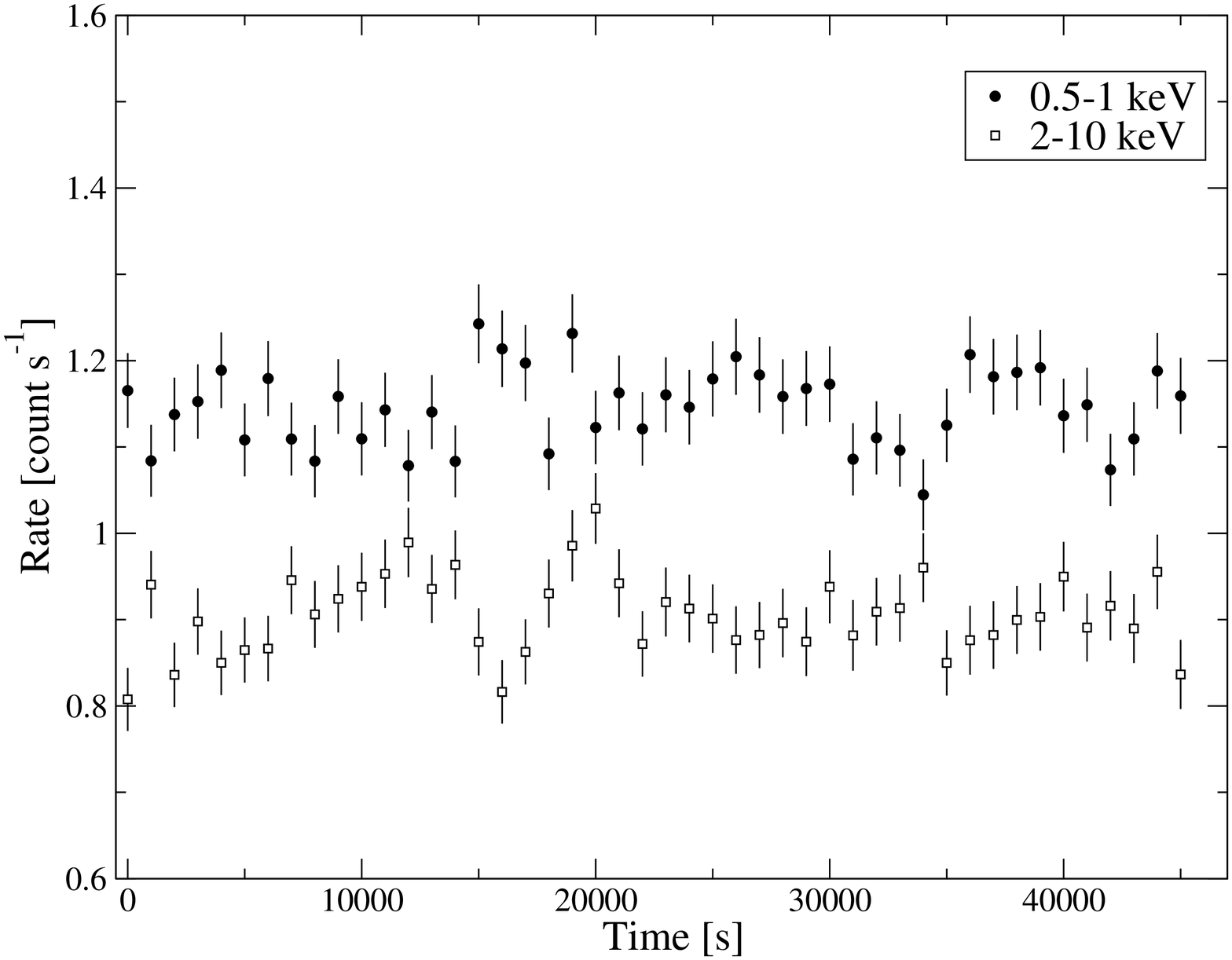}}
\caption{Soft (filled circles) and hard (open squares) \pn light curves of \eso
 (upper panel) and \cts (lower panel) binned by 1000\,s.}
 \label{4obj:eso-cts-light-curves}
\end{figure}

We analyzed the \pn\ soft (0.5-1\,keV) and hard (2-10\,keV) X-ray
background-subtracted light curves to investigate the
variability of the four observed sources during their {\em XMM-Newton}
observations. 
The overall behavior of the light curves of \eso and \cts seems to show a
constant average flux value and short-time, low-amplitude flux variations.
In the case of the \eso light curves (upper panel of 
Fig.\ \ref{4obj:eso-cts-light-curves}), the ratio between the maximum and
minimum count rates is 1.5$^{+0.3}_{-0.2}$ and 1.3$^{+0.3}_{-0.2}$ for the
soft and hard 
bands, respectively, thereby indicating no statistically significant variations.

For \cts light curves (lower panel of Fig.\ \ref{4obj:eso-cts-light-curves}),
the ratios of the maximum to minimum rate are 
1.3$\pm$0.1 and 1.4$\pm$0.2 for soft and hard bands, respectively. 
Therefore, taking the typical flux variations of this kind of
objects into account, the flux variation during the observation was of small
amplitude although significant ($3\sigma$) in the soft band, and only
marginally significant ($2\sigma$) in the hard band. 
Under a careful visual inspection of the light curves, there 
seems to be an anti-correlation between the variations in the soft and hard
light curves: an increment in flux in the soft band seems to
correspond to a decrement in the hard band flux, and {\em viceversa}. The
cross-correlation function between these curves does not suggest that this
behavior could be explained by a temporal delay. 
The soft vs.\ hard count rates representation shows a 
weak negative regression that could indicate an
anti-correlation. However, there is a large dispersion in the data, so we
cannot assert the presence of that anti-correlation based only on the 
present observations. 

\begin{figure}
\centering
\resizebox{\hsize}{!}{\includegraphics{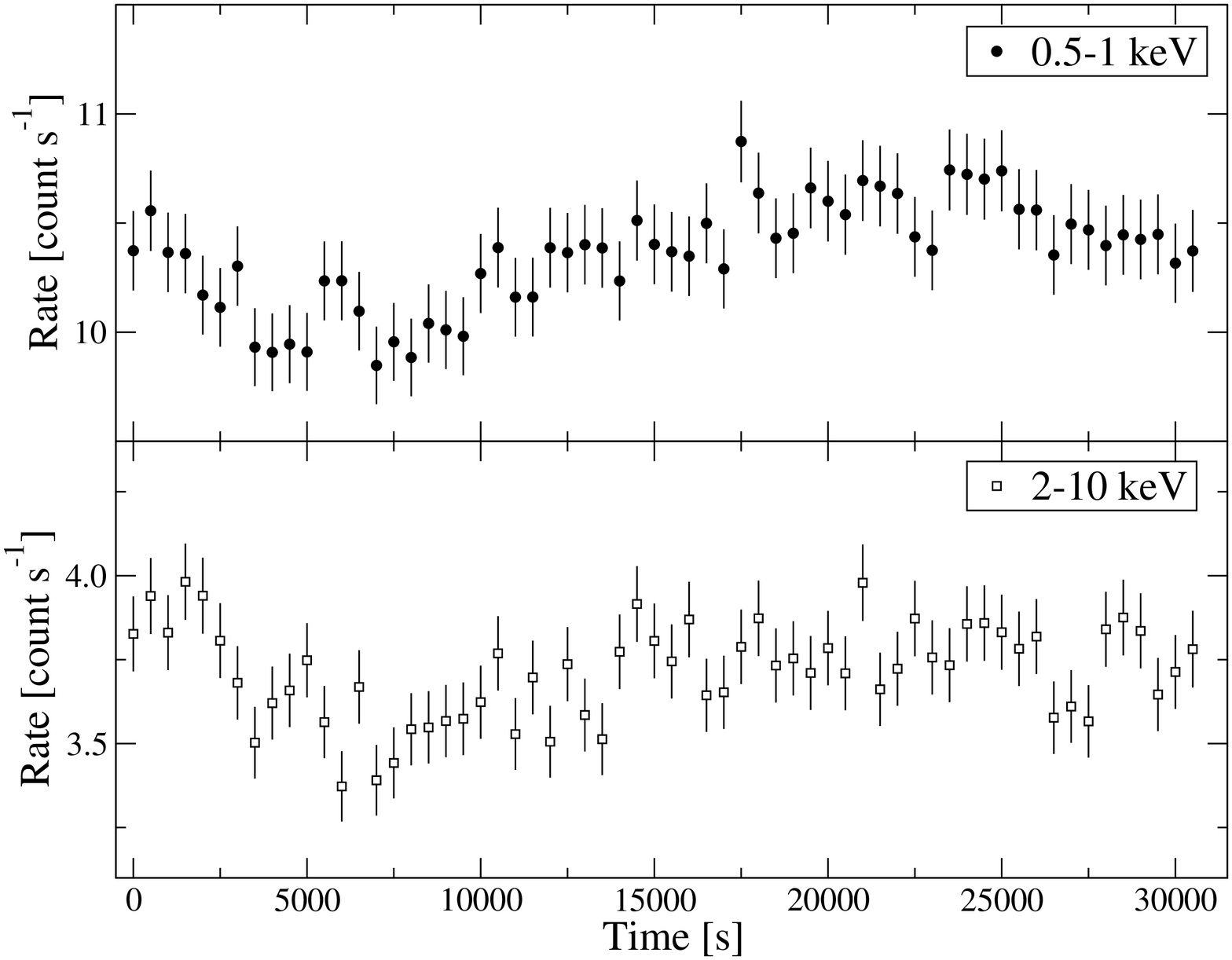}}\\
\vspace*{0.3cm}
\resizebox{\hsize}{!}{\includegraphics{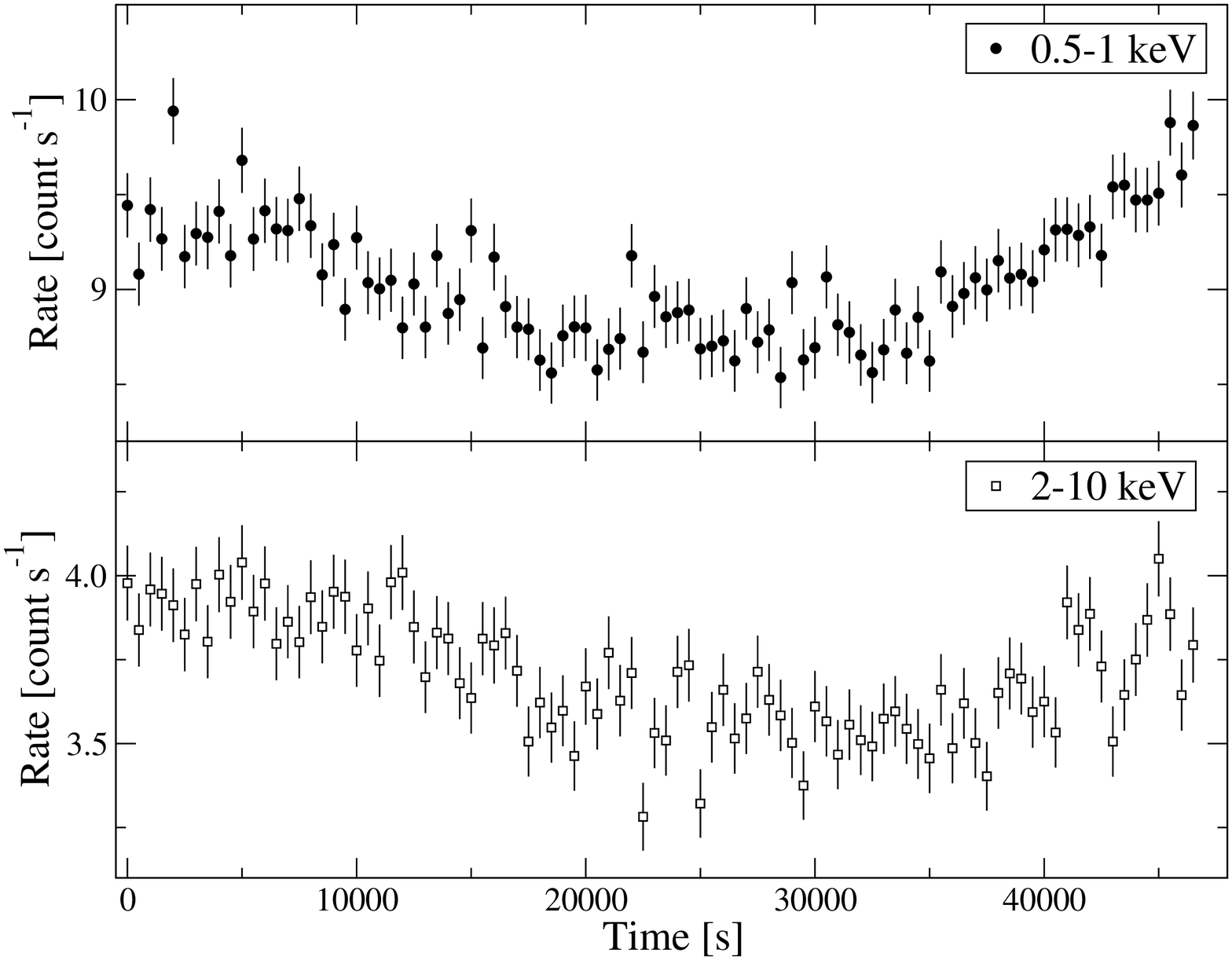}}\\
\caption{Soft (filled circles) and hard (open squares) \pn light curves of \he
  (upper panel) and \mkn (lower panel) binned by 500\,s.}
 \label{4obj:he-mkn-ligth-curves}
\end{figure}

The light curves of \he and \mkn are not well modeled by a constant flux
function. In the case of \he the ratios 
between the maximum and minimum count rates are 1.10$\pm$0.04 and
1.18$\pm$0.07, indicating that the flux variations are small but significant at
the $\sim2.5\sigma$ level (Fig. \ref{4obj:he-mkn-ligth-curves}, upper
panel). Flux variations in the two bands are well correlated without any
temporal delay.  
For \mkn the ratio between the maximum and minimum count rates for the
soft and hard light curves (1.16$\pm$0.04 and 1.23$\pm$0.07) indicates that
the amplitudes of the flux variations are small but significant at $4\sigma$
and $3\sigma$ levels, respectively (Fig. \ref{4obj:he-mkn-ligth-curves}, lower panel). 
In this case, there is a time lag of about 1000\,s 
between the flux variations in the soft and hard bands, indicating that
variations are seen first in the soft band and about 17 minutes later in the
hard band.

\begin{table*}[ht]
\begin{minipage}[t]{\textwidth}
\begin{center}
\caption{Comparison between X-ray fluxes from the literature and from
  this work.}
\label{4obj:x-flux}
\begin{tabular}{@{}llccccc@{}}
\hline\hline
\noalign{\smallskip}
Data & Obs.\ date & 0.2-2\,keV & 0.5-2\,keV & 0.1-2.4\,keV & 2-10\,keV & Ref. \\
\noalign{\smallskip}
\hline
\noalign{\smallskip}
\multicolumn{7}{c}{\bf \eso}  \\            
ROSAT  & 1990 Aug.\    & --                       & --  & 48.3           & -- & 1 \\
ROSAT  & 1990 Aug.\    & 21.9                     & --  &  --            & -- & 2 \\
XMM    & 2004 Mar.\    & $2.39\pm 0.07$ & $1.73\pm 0.05$  & $2.71\pm 0.09$ & $3.1\pm 0.1$ & 3 \\
SWIFT  & 2008 Oct.\ 12 &  2.8                     & --  &  --            & -- & 4 \\
SWIFT  & 2008 Oct.\ 26 &  2.0                     & --  &  --            & -- & 4 \\[1pt]   
\noalign{\smallskip}
\hline
\noalign{\smallskip}
\multicolumn{7}{c}{\bf \he}  \\             
ROSAT  & 1990 Dec.\    & --                       & --  & 75.8           & --   & 1 \\
XMM    & 2004 Jun.\    & --                       & --  &  --            & 28.3 & 5 \\[1pt]
XMM    & 2004 Jun.\    & $37.69\pm 0.05$ & $25.26\pm 0.05$  & $40.72\pm 0.07$
& ${29_{-3}^{+1}}$ & 3\\                                                     
SWIFT  & 2006 Jul.\    & --                       & 10.0&  --            & 14.3 & 6 \\
\noalign{\smallskip}
\hline
\noalign{\smallskip}
\multicolumn{7}{c}{\bf \cts}  \\            
ROSAT  & 1990 Oct.\    & --                       & --  & 44.4           & --   & 1 \\  
ASCA   & 1998 May      & --                       & 14.0&    --          & 29.0 & 7 \\ [1pt]
XMM    & 2004 Oct.\    & $4.07\pm 0.05$ & $3.31\pm 0.05$  & $4.66\pm 0.06$ & $7.6\pm 0.1$ & 3 \\
\noalign{\smallskip}
\hline
\noalign{\smallskip}
\multicolumn{7}{c}{\bf \mkn}  \\            
ROSAT  & 1990 Oct.\    & --                       & --  & 23.0           & --  & 1 \\
ROSAT  & 1990 Oct.\    & 13.8                     & --  &    --          & --  & 2 \\
ROSAT  & 1991 Oct.\    & 55.0                     & --  &    --          & --  & 2 \\
ASCA   & 1998 May      & --                       & 16.0&    --          & 28.0& 7 \\
XMM    & 2004 Nov.\    & --                       & --  &    --          & 28.5& 5 \\
XMM    & 2004 Nov.\    & --                       & --  &    --          & 28  & 8 \\[1pt]
XMM    & 2004 Nov.\    & $35\pm 3$ & $22\pm 3$  & $38\pm 4$ & $29\pm 2$ & 3 \\
SWIFT  & 2010 Jan.\ 06 & 51.3                     & --  &    --          & --  & 4 \\
SWIFT  & 2010 Jan.\ 12 & 52.4                     & --  &    --          & --  & 4 \\
\noalign{\smallskip}
\hline
\noalign{\smallskip}
\end{tabular}           
\end{center}
\end{minipage} \par           
Fluxes are given in $10^{-12}$\,\funits. Errors quoted are at 90\% confidence level. 
References: (1) \citeauthor{1999A&A...349..389Voges} (1999; Flux1);
(2) \cite{2001A&A...367..470Grupe}; 
(3) This work; 
(4) \cite{2010ApJS..187...64Grupe}; 
(5) \cite{2007MNRAS.382..194Nandra}; 
(6) \cite{2009ApJ...690.1322Winter}; 
(7) from the TARTARUS database (http://tartarus.gsfc.nasa.gov/); 
(8) \cite{2009A&A...507..159DeMarco}.
\end{table*}

The X-ray fluxes available from the literature for these objects are
summarized in Table \ref{4obj:x-flux}. For this table we have taken the fluxes
listed as Flux1 in the {\em ROSAT} All-Sky Bright Source Catalog
\citep{1999A&A...349..389Voges} since it is obtained assuming a power law model
($\alpha_x=1.3$) with an absorbed column density fixed at the Galactic value.
Fitting also an absorbed power law to the same {\em ROSAT} All-Sky Survey
(RASS) data of \eso and \mkn, \cite{2004AJ....127..156Grupe} found very
different flux values (lower than those in \citealt{1999A&A...349..389Voges} by
factors of about 2.2 and 1.7, respectively) in almost the same interval. 
All these objects show a variation of large amplitude in its soft X-ray flux,
from a factor of about 2 for \he to a factor of about 18 for \eso.  
Fluxes in the 0.1-2.4\,keV range taken from {\em ROSAT} observations 14 years
before the {\em XMM-Newton} ones, are found to be larger than measured by us
by about one order of magnitude for \eso and \cts and by only a factor
of 2 for \he. However, in the case of \mkn, our estimation of this flux is
larger than the one obtained from {\em ROSAT} by a factor of $1.7$. 
For this object 
the maximum soft X-ray flux reported in the literature was measured
one year later (1991), also with {\em ROSAT}, and is a factor $\sim1.6$ higher
than our {\em XMM-Newton} value.
 
{\em SWIFT} observations exist for \eso, \he, and \mkn. For the first, similar
values to those derived by us are reported.
In the case of \he and \mkn our results are higher and lower than those
derived from the {\em SWIFT} data by factors of 2.5 and 1.5,
respectively. \mkn shows very similar flux values between the data acquired in
1991 and 2010 by {\em ROSAT} and {\em SWIFT}, respectively.

In the 2-10\,keV range the amplitude of the flux variations is larger than a
factor of 2 for \he and 3.8 for \cts. For \eso there are no published
data in this spectral range, and the hard {\em XMM-Newton} (2004) and 
{\em ASCA} (1998) values for \mkn
are very similar, although the {\em ASCA} soft flux is a factor 
$\sim1.4$ lower than the soft {\em XMM-Newton} flux. Unfortunately, for the
galaxies studied here there is no data in the hard spectral range taken
contemporaneously to the observed highest soft X-ray flux. 
Taking the fluxes in Table\,\ref{4obj:x-flux} into account for these objects,
it becomes clear that, when they 
were observed by the {\em XMM-Newton} satellite, only \cts was in the lowest 
reported activity state, 
\eso was almost in the lowest reported state (very similar to {\em SWIFT}
observations), and \he and \mkn were in an intermediate state.


\section{X-ray spectral analysis \label{sec:x-rays}}

All the EPIC data were checked for no pile-up effects using the SAS task 
{\sc epatplot}. We have found these effects to be present in the \mos1 and
\mos2 spectra of \he and \mkn. 
For these objects we excised the core of the source (i.e., the core
of the point spread function) using an annular shape for the source extraction
region of the MOS data. The inner radii were selected as the minimum radii
that ensures a negligible fraction (less than 5\%) of pile-up in the data. 
The EPIC spectra were extracted using standard parameters. Except for those
data affected by pile-up, the source extraction areas were circular regions. 
All the EPIC spectra were binned to have at least 30 counts per bin. 

The RGS data were processed with default parameters except that we
discarded potentially problematic cold pixels. All the RGS spectra were
binned, losing spectral resolution but increasing the signal-to-noise ratio. 
We selected a geometrical binning since this method
minimizes any smoothing of the absorption and emission features.
We binned the RGS data of \eso, \he, and \cts using 15 channels per bin, 
while 10 channels per bin were used in the case of \mkn.
Because the default RGS spectral bin size is
10\,m\AA\ at 15\,\AA, our final spectra for \eso, \he, and \cts have bins of
about 150\,m\AA\ at that wavelength, while those of \mkn have bins of about
100\,m\AA.

All spectra were fitted using {\em Sherpa} package of {\em CIAO 3.4}
\citep{2001SPIE.4477...76Freeman,2006SPIE.6270E....Silva}. 
We used the $\chi^2$ statistics with the Gehrels variance function
\citep{1986ApJ...303..336Gehrels} and the Powell optimization method, 
the first because it is based on Poisson statistics for a small number of
counts in a bin and on Binomial statistics otherwise, and the second because
it is a robust direction-set method for finding the nearby fit-statistical
minimum.


\subsection{Low-resolution spectra \label{lr-spectra}}

\begin{table*}[ht]
\begin{minipage}[t]{\textwidth}
\begin{center}
\caption[Best-fit model to the continuum of the EPIC spectra in the
0.35-10\,keV range.]{Best-fit model to the continuum of the EPIC spectra in
  the 0.35-10\,keV range.}
\label{4obj:epic-continuum}
\begin{tabular}{@{}l ll @{}l ll @{}l ll l@{}}
\hline\hline
\noalign{\smallskip}				  
   & \multicolumn{2}{c}{Hard power law} 
  && \multicolumn{2}{c}{Black body} 
  && \multicolumn{2}{c}{Soft power law}
  & \multicolumn{1}{c}{$\chi^{2}_{\nu}$/$\mathit{dof}$}   \\ 
\cline{2-3} \cline{5-6} \cline{8-9}

 & \multicolumn{1}{c}{$\Gamma$} & \multicolumn{1}{c}{$K_{pwlw}$}   
&& \multicolumn{1}{c}{kT} & \multicolumn{1}{c}{$K_{bb}$} 
&& \multicolumn{1}{c}{$\Gamma$} & \multicolumn{1}{c}{$K_{pwlw}$} \\
\noalign{\smallskip}
\hline
\noalign{\smallskip}
\eso & $1.69_{-0.04}^{+0.04}$ & $7.1_{-0.3}^{+0.3}$ 
     && $0.16_{-0.01}^{+0.01}$ & $7.1_{-0.3}^{+0.3}$
     && \multicolumn{1}{c}{---} & \multicolumn{1}{c}{---}
     & 0.76/716\\[3pt]
\he  & $1.19_{-0.07}^{+0.07}$ & $19_{-3}^{+3}$ 
     && \multicolumn{1}{c}{---} & \multicolumn{1}{c}{---}
     && $2.61_{-0.03}^{+0.03}$ & $102_{-3}^{+3}$ 
     & 0.90/1914\\[3pt]
\cts & $1.55_{-0.02}^{+0.02}$ & $14.5_{-0.2}^{+0.2}$ 
     && $0.131_{-0.004}^{+0.004}$ & $1.46_{-0.09}^{+0.09}$
     && \multicolumn{1}{c}{---} & \multicolumn{1}{c}{---}
     & 0.88/1486\\[3pt]
\mkn & $1.3_{-0.1}^{+0.2}$ & $24_{-5}^{+21}$ 
     && \multicolumn{1}{c}{---} & \multicolumn{1}{c}{---}
     && $2.5_{-0.1}^{+0.3}$ & $78_{-12}^{+5}$ 
     & 0.90/2187\\[3pt]
\noalign{\smallskip}
\hline
\noalign{\smallskip}
\end{tabular}
\end{center}
\end{minipage}\par
Power-law normalizations ($K_{pwlw}$) in units of $10^{-4}$
ph\,\,keV$^{-1}$\,cm$^{-2}$\,s$^{-1}$ at 1 keV; kT in keV; $K_{bb}$ in
$10^{-5}$\,$L_{39}/D_{10}^{2}$ where $L_{39}$ is the source luminosity in
units of $10^{39}$ erg\,s$^{-1}$ and $D_{10}$ is the distance to the source in
units of 10 kpc. The Galactic $N_\mathrm{H}$
values are fixed (see Table \ref{objects}). Errors quoted are at 90\%
confidence level. 
\end{table*}

For each source we start by fitting only the \pn spectrum in the hard band
(2.0-10.0~keV), then we check for soft excess extending the hard band model to
the soft band, and finally we fit the three EPIC (pn, MOS\,1 and MOS\,2)
spectra simultaneously in the whole range considered (0.35-10\,keV). 
Below 1 keV, all the objects analyzed in this work show an X-rays excess over
the extrapolation of the harder power law flux.
Soft excess components are modeled in the simplest way. We first use a black
body component, and in the case this model does not represent the observed
excess, we use a power law. 
Because the Galactic absorption has to be taken into account, we include,
in all the source models mentioned from here on, a component that models the
Galactic neutral absorption with the hydrogen column density ($N_{\mathrm H}$)
fixed to the values listed in Table \ref{objects}.
In Table \ref{4obj:epic-continuum} we list -for each source- the best-fit
model continuum parameters found by performing the simultaneous EPIC fit. 

The criteria we have used to include Gaussian profiles in the models
is the same as described in \cite{2009A&A...505..541Cardaci}.
The lines are added constraining their energies to vary in a small range
around their laboratory energies ($E_{lab}$) and taking the
redshift of the source into account. Since we cannot use only the
F-test as a reliable criterion to compute the statistical significance
of a Gaussian line \citep{2002ApJ...571..545Protassov}, we use a combined
method to decide whether to include a line in our final
model. 
Along with the F-test, for which the minimum significance level considered 
is 95\%, 
we check that the residuals of the new fit at the position of the
tested line is less than 1 sigma, 
and finally we also ensure that the wavelength positioning of the line
be consistent with its laboratory wavelength (taking the galaxy
redshift into account).

Except for \mkn, we find that the fluorescent Fe-K$\alpha$ emission line is
significant to the fit (with more than 99\% significance). 
The Fe-K$\alpha$ line widths, $\sigma$, are small
enough to be considered equal to the instrumental resolution.

After analyzing the residuals of the \eso fit, we find a second narrow line
at about 7\,keV that could be originated by Fe{\sc xxvi}
(E$_{\mathrm{lab}}=6.966$\,\kev). The line width has been fixed to the
instrumental resolution. This line is statistically significant to the
fit (with more than 95\% significance).

In the residuals of the \he fit, we see a wide 4$\sigma$ Gaussian
shape excess between 0.4 and 0.8\,\kev. This excess is well described with a 
Gaussian line centered at $0.579_{-0.007}^{+0.006}$\,\kev 
($\sim21.4$\,\AA) with a dispersion of
$0.06_{-0.01}^{+0.01}$\,\kev ($28000_{-5000}^{+5000}$\,\kms). It could be
interpreted as a blend of O{\sc vii} lines. 

On the \cts fit, there is still a small (2 sigma)
line-like residual around 6\,\kev that might indicate a relativistic
shape for the Fe-K$\alpha$ line. Unfortunately, 
the signal-to-noise ratio of this line is low, and with these data we cannot
test the hypothesis of a potential relativistic shape of the 
Fe-K$\alpha$ line any further.

In the case of \mkn, although there are residuals around 6\,\kev that could
be interpreted as signatures of an Fe-K$\alpha$ emission line, including
this component (with only 70\% significance) proved not to be
significant to the fit. After the modeling of 
the soft excess component still remains a broad excess in the residuals
between 0.5 and 0.6\,\kev. We included three Gaussian
components at the energy position of the O{\sc vii} triplet emission lines
(0.5740, 0.5686, 0.5610\,\kev) to model it.
These lines were included with the relative energies between them fixed to the
theoretical values, and having the same dispersion. 
Since the normalizations of two of the lines take lower
values (around $2\times10^{-5}$ and 
$3\times10^{-6}$\,ph\,\,cm$^{-2}$\,s$^{-1}$) than the third line
(about $4\times10^{-4}$\,ph\,\,cm$^{-2}$\,s$^{-1}$) this
triplet is dominated by only one component.

The parameter values of the final EPIC model of each object are
listed in Tables \ref{4obj:epic-continuum} and \ref{4obj:epic-lines}.

\begin{table}[ht]
\begin{minipage}[t]{\columnwidth}
\begin{center}
\caption{Line parameters included in the best-fit model to the
simultaneous three EPIC spectra in the 0.35-10\,keV range.} 
\label{4obj:epic-lines}
\begin{tabular}{@{}l @{\hspace{2mm}} l l lll @{}}
\hline\hline
\noalign{\smallskip}				  
Object & line &\multicolumn{1}{c}{$E_{\mathrm{lab}}$} &\multicolumn{1}{c}{$E_{\mathrm{rest}}$} & \multicolumn{1}{c}{$\sigma$} & \multicolumn{1}{c}{$K_{\mathrm{line}}$}\\
\noalign{\smallskip}
\hline
\noalign{\smallskip}
\eso & Fe K$\alpha$  & 6.4 & $6.38_{-0.05}^{+0.06}$ & $0.004^{a}$ & $0.7_{-0.2}^{+0.2}$ \\[3pt]
     & Fe{\sc xxvi}  & 6.966 & $6.97^{a}$ & $0^{b}$ & $0.5_{-0.2}^{+0.2}$ \\[3pt]
\he  & Fe K$\alpha$  & 6.4 & $6.39_{-0.04}^{+0.04}$ & $0.07_{-0.04}^{+0.05}$ & $1.8_{-0.6}^{+0.7}$ \\[3pt]
     & O{\sc vii}    & blend &$0.579_{-0.007}^{+0.006}$ & $0.06_{-0.01}^{+0.01}$ & $90_{-10}^{+10}$ \\[3pt]
\cts & Fe K$\alpha$  & 6.4 & $6.39_{-0.03}^{+0.02}$ & $0.06_{-0.04}^{+0.04}$ & $1.2_{-0.3}^{+0.3}$ \\[3pt]
\mkn & O{\sc vii} (r)& 0.5740 & $0.59_{-0.17}^{+0.12}$ & $0.05_{-0.04}^{+0.03}$ & $2.2^{a}$ \\[3pt]
     & O{\sc vii} (i)& 0.5686 & $0.5871^{c}$ & $0.05^{c}$ & $0.3^{a}$ \\[3pt]
     & O{\sc vii} (f)& 0.5610 & $0.5795^{c}$ & $0.05^{c}$ & $44^{a}$ \\[3pt]
\noalign{\smallskip}
\hline
\noalign{\smallskip}
\end{tabular}
\end{center}
\end{minipage}
$^a$Unconstrained. $^b$Fixed to the instrumental resolution. 
$^c$Fixed to the energy and width of the recombination component.  
$E_{\mathrm{rest}}$ (line energies in the rest frame of the source),
$E_{\mathrm{lab}}$ and $\sigma$ of the emission lines are given in keV;
and line normalizations $K_\mathrm{line}$ in units of $10^{-5}$ ph\,\,cm$^{-2}$\,s$^{-1}$. 
Errors quoted are at 90\% confidence level.
\end{table}


\subsection{High-resolution X-ray spectra \label{hr-spectra}}

The EPIC spectra, with their broad energy coverage provide 
a good picture of the shape of the continuum emission of the source and its
wide emission features. 
However, the search for narrow features requires using the higher resolution
spectra provided by the RGS instruments. 
On the other hand, by fitting only the RGS spectra, one could find solutions
not completely compatible with those derived from the analysis of the EPIC
spectra alone. Therefore a comprehensive five-spectra fit including the three
EPIC and the two RGS spectra has been made.
The \pn\ and \mos\ data are restricted to the 0.35\,--\,10\,keV energy range,
while the \rgs\ data are taken in the 0.41 and 1.8\,keV (about 6.9-30.2\,\AA)
range.

\begin{table*}[ht]
\begin{minipage}[t]{\textwidth}
\begin{center}
\caption{Continuum parameters of the comprehensive \pn, \mos, and \rgs best-fit model for \eso, \he, \cts, and \mkn.} 
\label{4obj:all-continuum}
\begin{tabular}{@{}l ll @{}l ll @{}l ll l@{}}
\hline\hline
\noalign{\smallskip}				  
   & \multicolumn{2}{c}{Hard power law} 
  && \multicolumn{2}{c}{Black body} 
  && \multicolumn{2}{c}{Soft power law}
  & \multicolumn{1}{c}{$\chi^{2}_{\nu}$/$\mathit{dof}$}   \\ 
\cline{2-3} \cline{5-6} \cline{8-9}
\noalign{\smallskip}
 & \multicolumn{1}{c}{$\Gamma$} & \multicolumn{1}{c}{$K_{pwlw}$}   
&& \multicolumn{1}{c}{kT} & \multicolumn{1}{c}{$K_{bb}$} 
&& \multicolumn{1}{c}{$\Gamma$} & \multicolumn{1}{c}{$K_{pwlw}$} \\
\noalign{\smallskip}
\hline
\noalign{\smallskip}
\eso  & $1.78_{-0.05}^{+0.04}$ & $7.9_{-0.4}^{+0.2}$ 
& & $0.08_{-0.04}^{+0.03}$ & $0.3_{-0.2}^{a}$ 
& & \multicolumn{1}{c}{---} & \multicolumn{1}{c}{---} 
& 0.78/921 \\[3pt]
\he   & $1.22_{-0.07}^{+0.07}$ & $20_{-3}^{+3}$ 
& & \multicolumn{1}{c}{---} & \multicolumn{1}{c}{---} 
& & $2.62_{-0.03}^{+0.03}$ & $101_{-3}^{+3}$ 
& 0.92/2097 \\[3pt]
\cts  & $1.56_{-0.02}^{+0.02}$ & $14.7_{-0.2}^{+0.2}$ 
& & $0.126_{-0.004}^{+0.004}$ & $1.43_{-0.09}^{+0.09}$ 
& & \multicolumn{1}{c}{---} & \multicolumn{1}{c}{---} 
& 0.91/1665 \\[3pt]
\mkn  & $1.34_{-0.07}^{+0.10}$ & $28_{-5}^{+9}$ 
& & \multicolumn{1}{c}{---} & \multicolumn{1}{c}{---} 
& & $2.59_{-0.09}^{+0.34}$ & $75_{-11}^{+4}$ 
& 0.89/2513 \\
\noalign{\smallskip}
\hline
\noalign{\smallskip}
\end{tabular}
\end{center}
\end{minipage}\par
$^a$Unconstrained. Power-law normalizations ($K_{pwlw}$) in units of $10^{-4}$
ph\,\,keV$^{-1}$\,cm$^{-2}$\,s$^{-1}$ at 1 keV; kT in keV; $K_{bb}$ in
$10^{-5}$\,$L_{39}/D_{10}^{2}$ where $L_{39}$ is the source luminosity in
units of $10^{39}$ erg\,s$^{-1}$ and $D_{10}$ is the distance to the source in
units of 10 kpc. Normalizations $K_{pwlw}$ and $K_{bb}$ correspond
to the \pn\ spectrum.
The galactic $N_\mathrm{H}$
values are fixed (see Table \ref{objects}). Errors quoted are at 90\%
confidence level. \\
\end{table*}

\begin{table}[ht]
\begin{minipage}[t]{\columnwidth}
\begin{center}
\caption{Line parameters included in the
comprehensive \pn, \mos, and \rgs best-fit model for \eso, \he, \cts, and \mkn.}
\label{4obj:all-lines}
\begin{tabular}{@{}l @{\hspace{2mm}}l@{\hspace{2mm}}l lll @{}}
\hline\hline
\noalign{\smallskip}				  
Objects & line & \multicolumn{1}{c}{$E_\mathrm{lab}$} & \multicolumn{1}{c}{$E_\mathrm{rest}$} & \multicolumn{1}{c}{$\sigma$} & \multicolumn{1}{c}{$K_\mathrm{line}$} \\
\noalign{\smallskip}
\hline
\noalign{\smallskip}
\eso  & Fe K$\alpha$  & 6.4   & 6.38$_{-0.05}^{+0.06}$ & 0.004$^a$ & $0.7_{-0.2}^{+0.2}$ \\[3pt]
      & Fe{\sc xxvi}  & 6.966 & $6.97^a$ & 0$^b$ & $0.5_{-0.2}^{+0.2}$ \\[3pt] 
\he   & Fe K$\alpha$  & 6.4   & $6.39_{-0.04}^{+0.04}$ & $0.07_{-0.04}^{+0.05}$ & $1.8_{-0.6}^{+0.7}$ \\[3pt]
      & O{\sc vii}    & blend & $0.578_{-0.007}^{+0.006}$ & $0.056_{-0.009}^{+0.010}$ & $90_{-10}^{+10}$ \\[3pt]
\cts  & Fe K$\alpha$  & 6.4   & $6.40_{-0.03}^{+0.02}$ & $0.06_{-0.04}^{+0.04}$ & $1.2_{-0.03}^{+0.03}$ \\[3pt]
      & O{\sc vii} (r)& 0.5740& $0.574_{-0.006}^{+0.003}$ & $0.005_{^a}^{+0.005}$ & $7_{-4}^{+4}$ \\[3pt]
\mkn  & Fe K$\alpha$  & 6.4   & $6.43^a$& 0$^b$ & $1.5_{-0.5}^{+0.4}$ \\[3pt]
      & O{\sc vii}(r) & 0.5740& $0.59_{-0.16}^{+0.04}$ & $0.04_{-0.02}^{+0.02}$ & $1.6^{a}$  \\[3pt]
      & O{\sc vii}(i) & 0.5686& $0.5872^{c}$ & $0.04^{c}$ & $0.9^{a}$  \\[3pt]\noalign{\smallskip}
      & O{\sc vii}(f) & 0.5610& $0.5796^{c}$ & $0.04^{c}$ & $40_{-30}^{a}$  \\[3pt]\noalign{\smallskip}
      & O{\sc vii}(r)$^{d}$ & 0.5740& $0.5669^{a}$ & 0$^b$ & $7_{-7}^{+13}$  \\[3pt]
      & O{\sc vii}(i)$^{d}$ & 0.5686& $0.5615^{c}$ & 0$^b$ & $20_{-20}^{+50}$  \\[3pt]
      & O{\sc vii}(f)$^{d}$ & 0.5610& $0.5539^{c}$ & 0$^b$ & $7_{-4}^{+8}$  \\[3pt]
      & O{\sc viii}$^{d}$   & 0.6536& $0.652_{-0.005}^{+0.004}$ & $0^{b}$ & $3_{-3}^{+5}$  \\[3pt]
\hline
\noalign{\smallskip}
\end{tabular}
\end{center}
\end{minipage}\par
$^a$Unconstrained. $^b$Fixed to the instrumental resolution. $^c$Linked to
have the same relative energy with respect to the recombination component than
the laboratory lines of the O{\sc vii}. $^{d}$Only in the RGS model. 
The line energies 
($E_\mathrm{lab}$ and $E_\mathrm{rest}$) and $\sigma$ of the emission lines are given in keV;
and line normalizations $K_\mathrm{line}$ in units of $10^{-5}$ ph\,\,cm$^{-2}$\,s$^{-1}$. 
Errors quoted are at 90\% confidence level.
\end{table}

The best-fitting model obtained using the low-resolution spectra is used as the
first approximation for the five-spectra fit.
No relevant intrinsic neutral absorption components associated with the
studied sources have been found.

\begin{figure}
\centering
\includegraphics[angle=0,width=1\columnwidth]{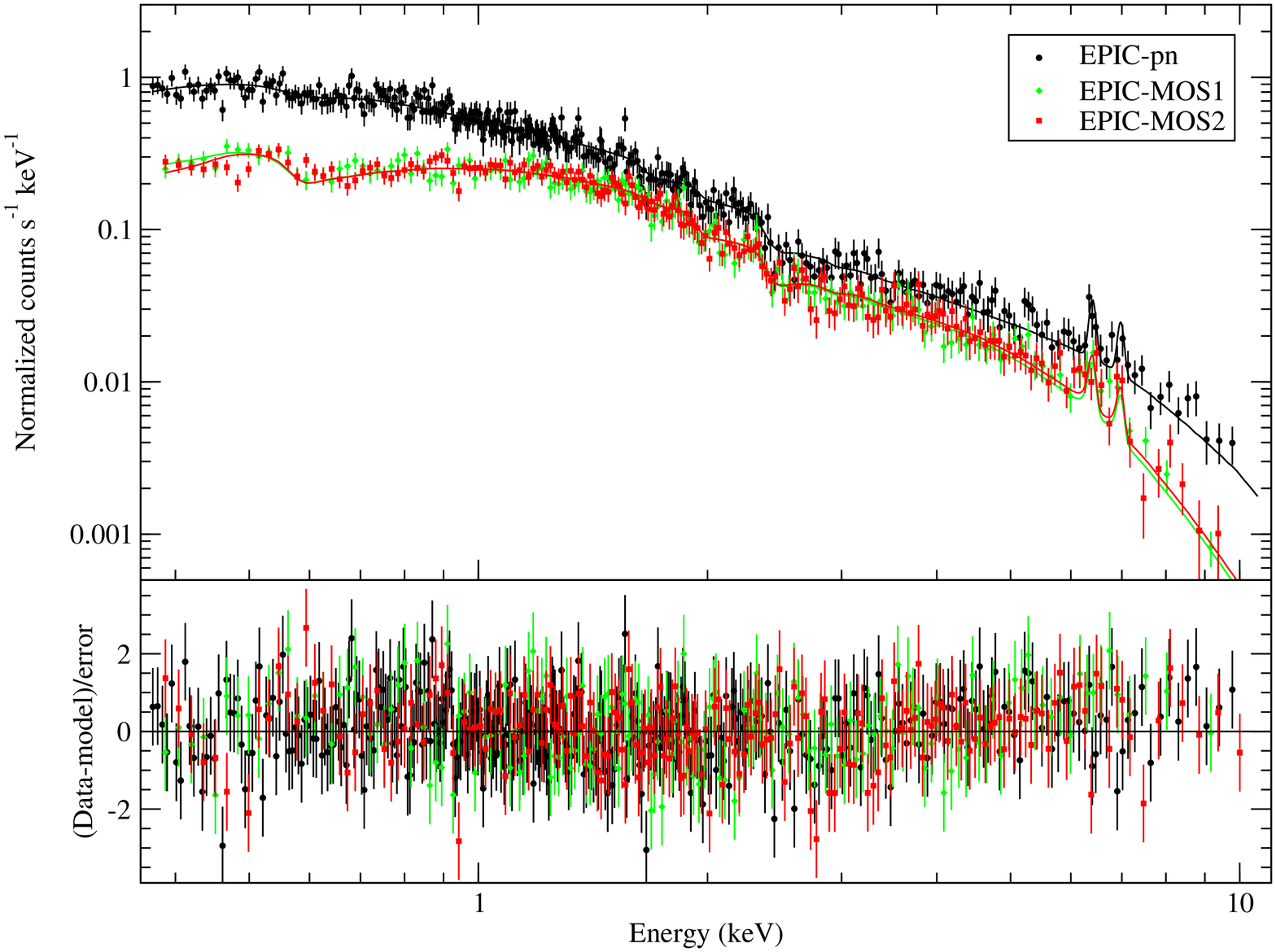}\\
\vspace*{0.3cm}
\includegraphics[angle=0,width=1\columnwidth]{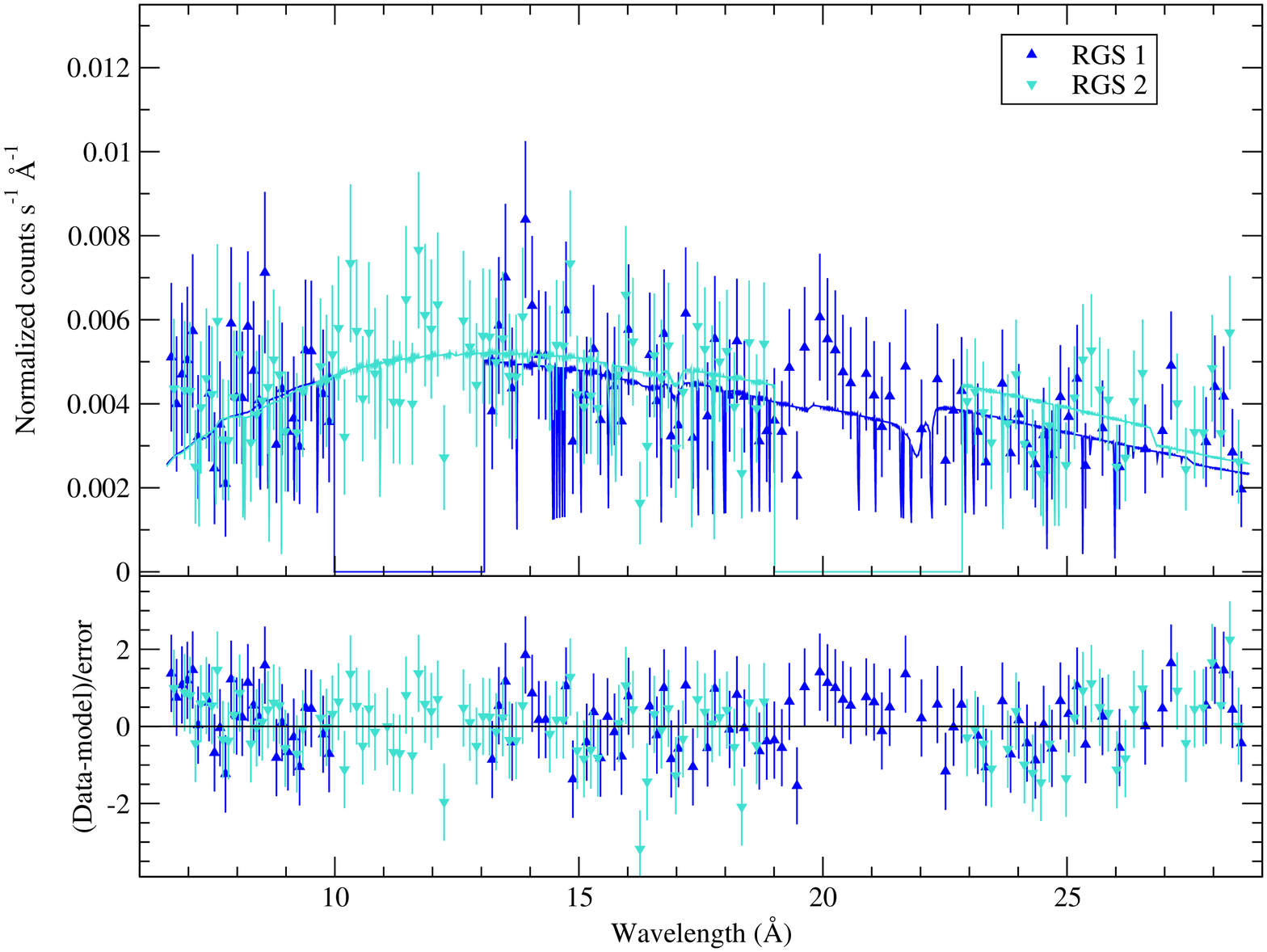}
\caption{\eso. Upper panel: Low-resolution X-ray spectra -in the rest
 frame- as obtained  with the \pn (black dots), \mos1 (green
 diamonds), and \mos2 (red squares) cameras. The best-fit model after a
 comprehensive fit of the \pn, \mos, and \rgs spectra (Tables
 \ref{4obj:all-continuum} and \ref{4obj:all-lines}) convolved with the
 instrument response of each camera is shown as a solid line.
 Lower panel: High-resolution X-ray spectra -in the rest frame- as
 obtained with the \rgs1 (blue up triangles), and \rgs2 (cyan down triangles),
 and  binned to 15 channels per bin. Solid lines are the convolution of the
 best-fit model after a comprehensive fit of the \pn, \mos, and \rgs spectra
 (Tables \ref{4obj:all-continuum} and \ref{4obj:all-lines}) with 
 instrument responses.}
 \label{eso:all-final}
\end{figure}

For the 
fit of \eso data we find a model with a
hard power law ($\Gamma=1.78_{-0.05}^{+0.03}$) slightly steeper than for the
simple EPIC fit and with a lower
black body temperature (kT$=0.08_{-0.04}^{+0.03}$). The parameter values of
the Fe lines identified on the EPIC spectra remain unchanged. 
In the residuals of the RGS fit (Fig.\ref{eso:all-final}), we notice a
wide feature located at 
about 20\,\AA. We tried to model it with a Gaussian component but we did not
find a satisfactory solution. 
The best-fitting model is plotted in Fig.\ \ref{eso:all-final}.

\begin{figure}
\centering
\includegraphics[angle=0,width=1\columnwidth]{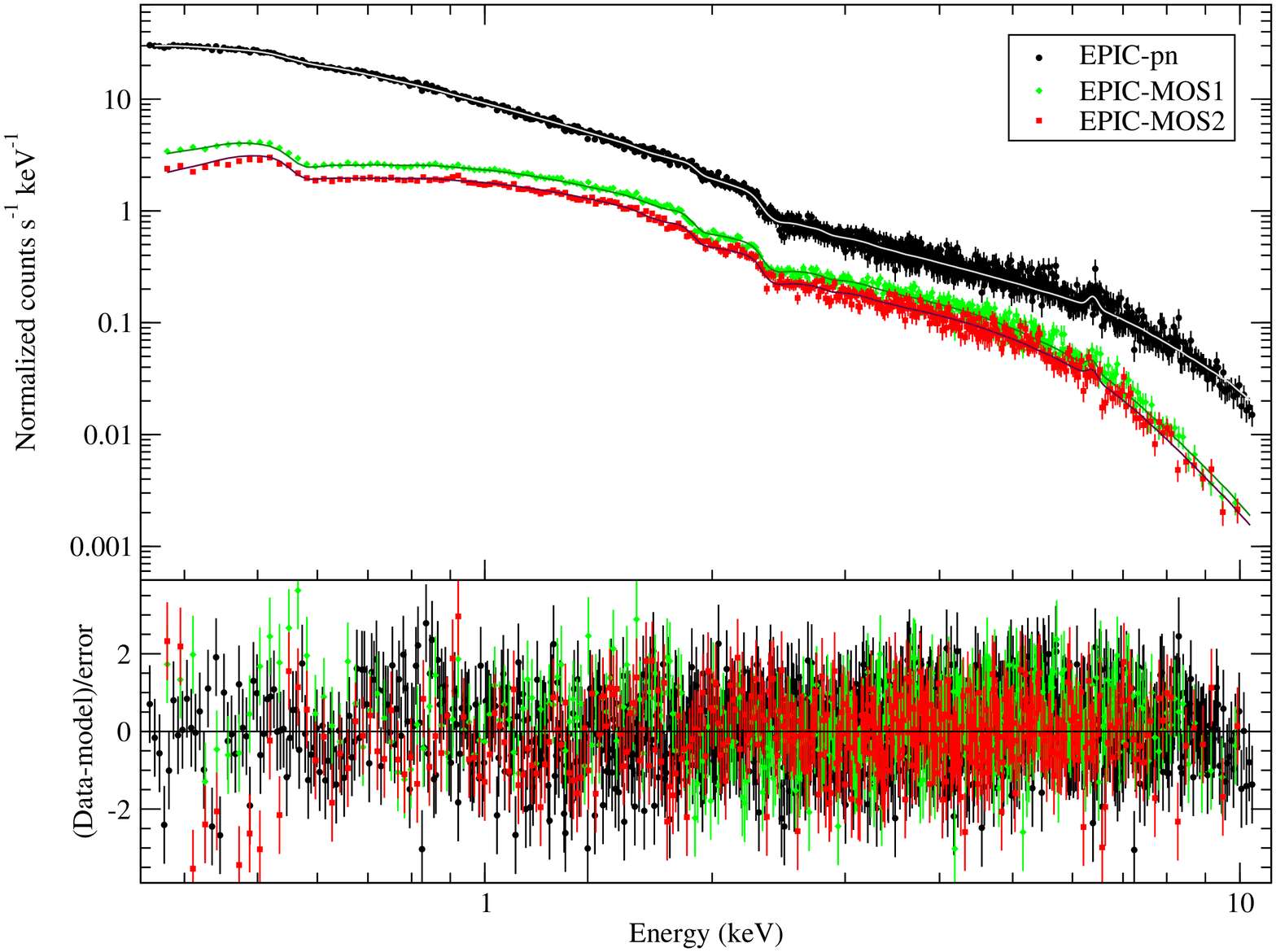}\\
\vspace*{0.3cm}
\includegraphics[angle=0,width=1\columnwidth]{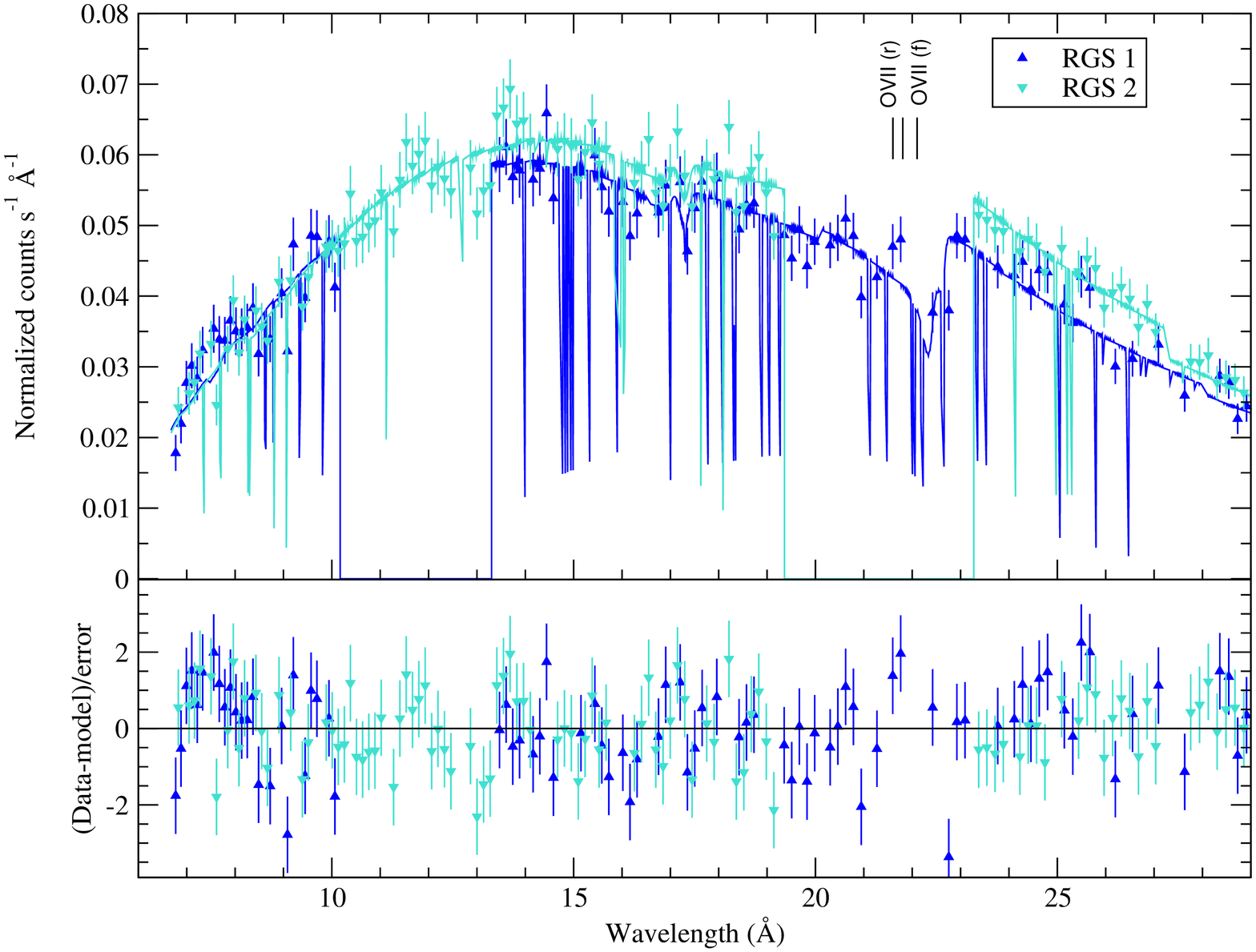}
\caption{As in Fig.\ \ref{eso:all-final} for \he.}
 \label{he:all-final}
\end{figure}

For \he, the same model that fits the \pn and \mos data is also
valid for describing the RGS data in the comprehensive fit
(\chired/\textit{dof}=0.92/2097). 
When substituting the broad oxygen line by three lines to account for the 
O{\sc vii} triplet, we find that the EPIC spectra are not well
modeled at soft energies. Residuals show about 4$\sigma$ departures from zero. 
We also tested for the presence of narrow O{\sc vii} triplet lines including
them only in the RGS spectra model but no satisfactory solution was found. 
Fig.\ \ref{he:all-final} shows the model and errors, together with the spectral
data. 

\begin{figure}
\centering
\includegraphics[angle=0,width=1\columnwidth]{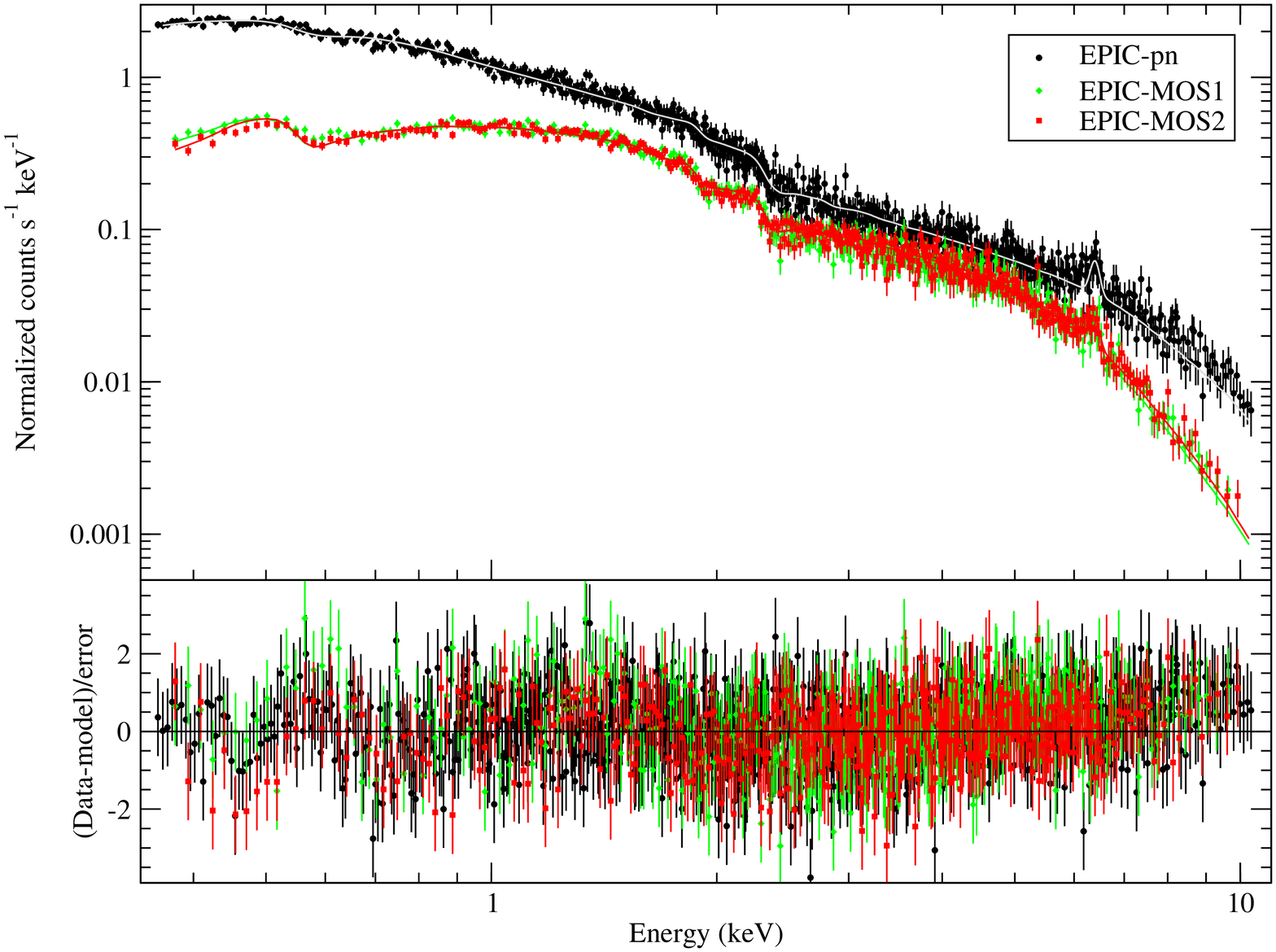}\\
\vspace*{0.3cm}
\includegraphics[angle=0,width=1\columnwidth]{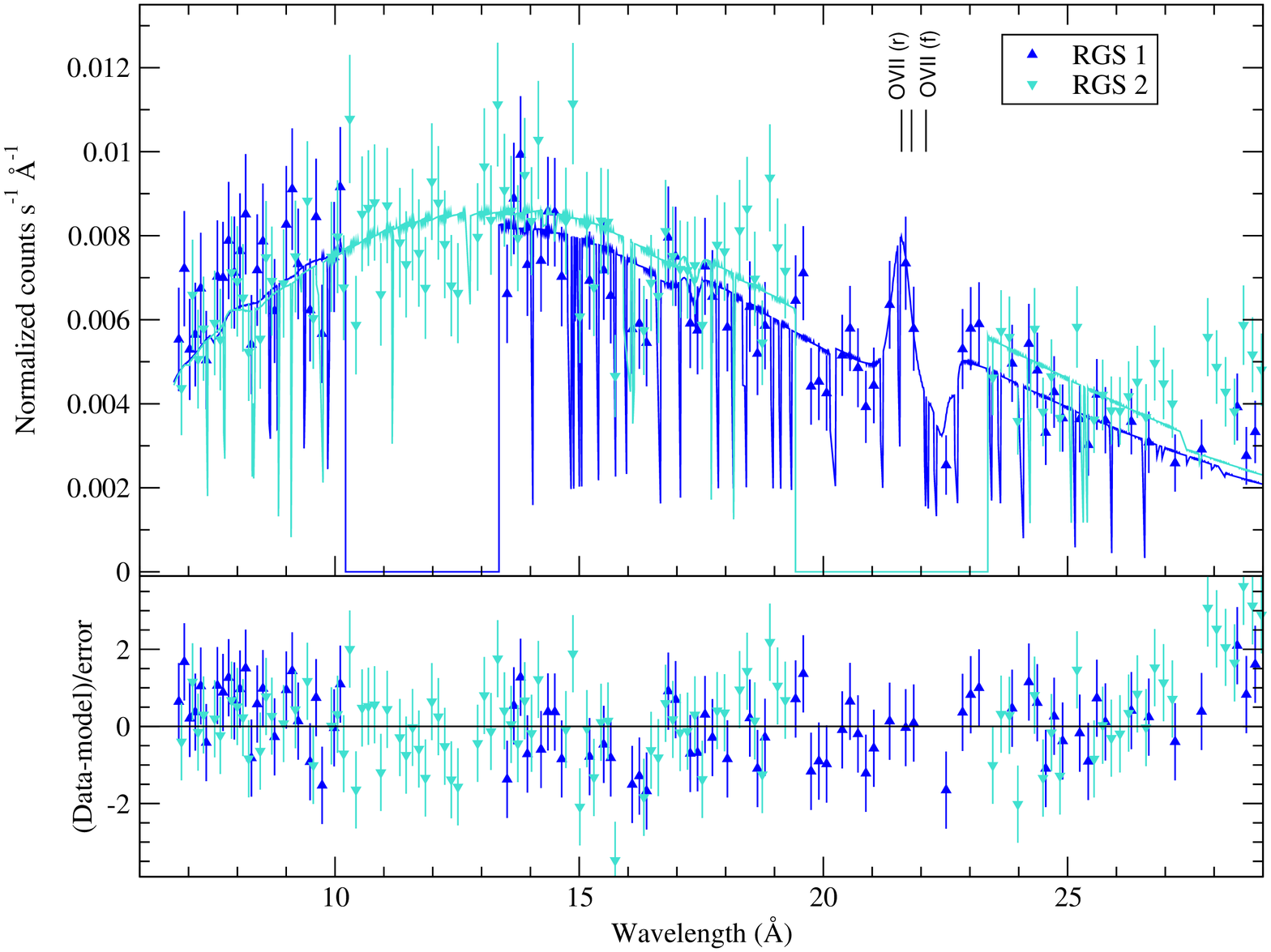}
\caption{As in Fig.\ \ref{eso:all-final} for \cts.}
 \label{cts:all-final}
\end{figure}

Also in the case of \cts, the same model provides a good fit to both the EPIC
and the RGS data 
However, the residuals in the high-resolution spectra fit
indicate the need to introduce narrow Gaussian lines at the theoretical
wavelengths of the O{\sc vii} and O{\sc viii} lines. 
Only one of the O{\sc vii} lines is found (the normalizations of the other two
are lower than $10^{-9}$\,\normunits), and it is significant to the fit (more
than 99\% confidence level). The O{\sc viii} emission line is not
statistically significant to the fit.
In this object, the wide feature in absorption that arises between 15 and
17\,\AA\ could be the signature of partially ionized material absorbing
the source radiation. We tested for the possible presence of such a component
using the PHASE photoionization code \citep{2003ApJ...597..832Krongold} and
find a hydrogen column density lower than $10^{20}$\,cm$^{-3}$ for
the absorbing gas. Such a component does not affect the X-ray
bands analyzed in the present work. 
The best-fitting model is plotted in Fig.\ \ref{cts:all-final}.

\begin{figure}
\centering
\includegraphics[angle=0,width=1\columnwidth]{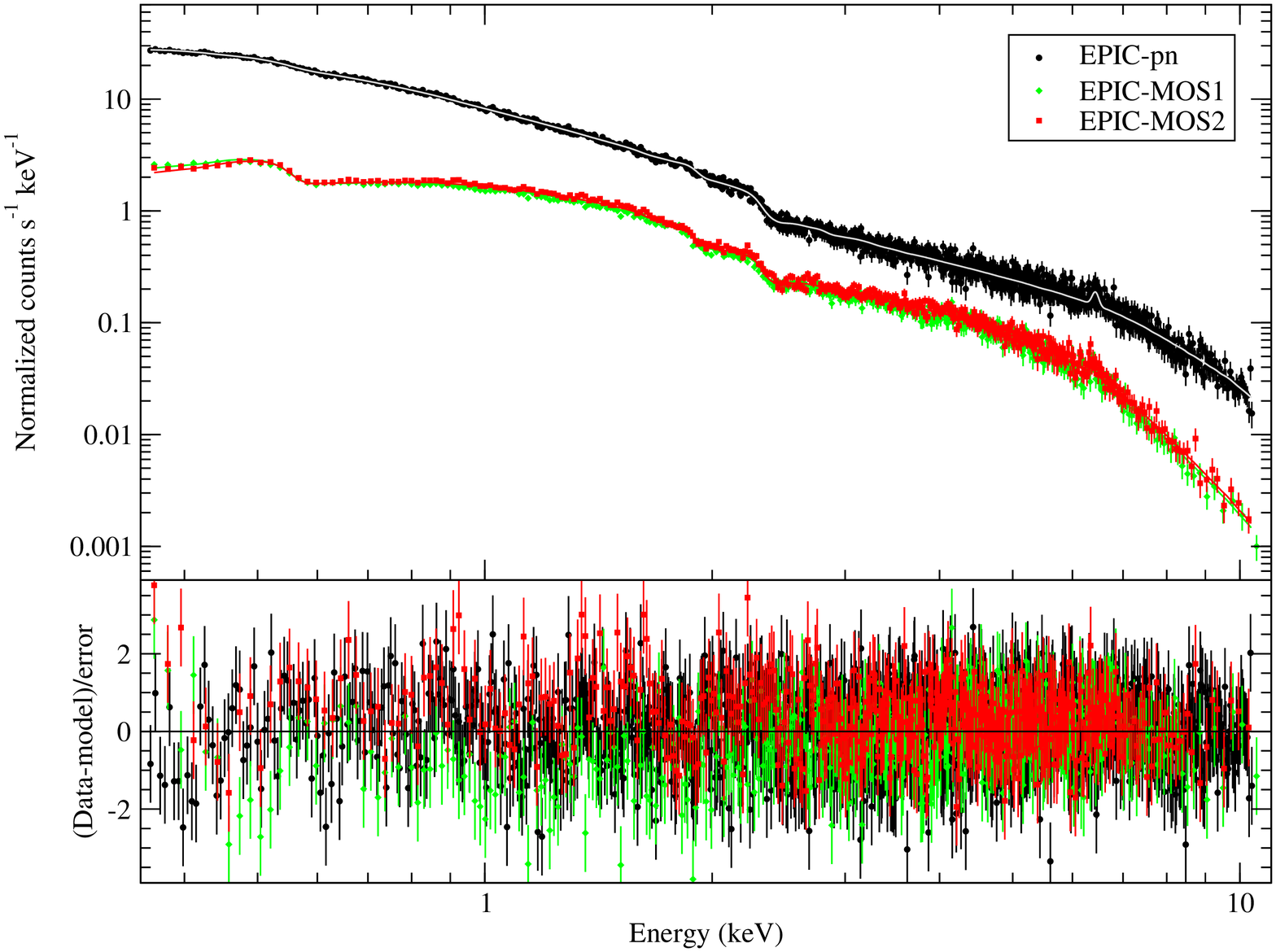}\\
\vspace*{0.3cm}
\includegraphics[angle=0,width=1\columnwidth]{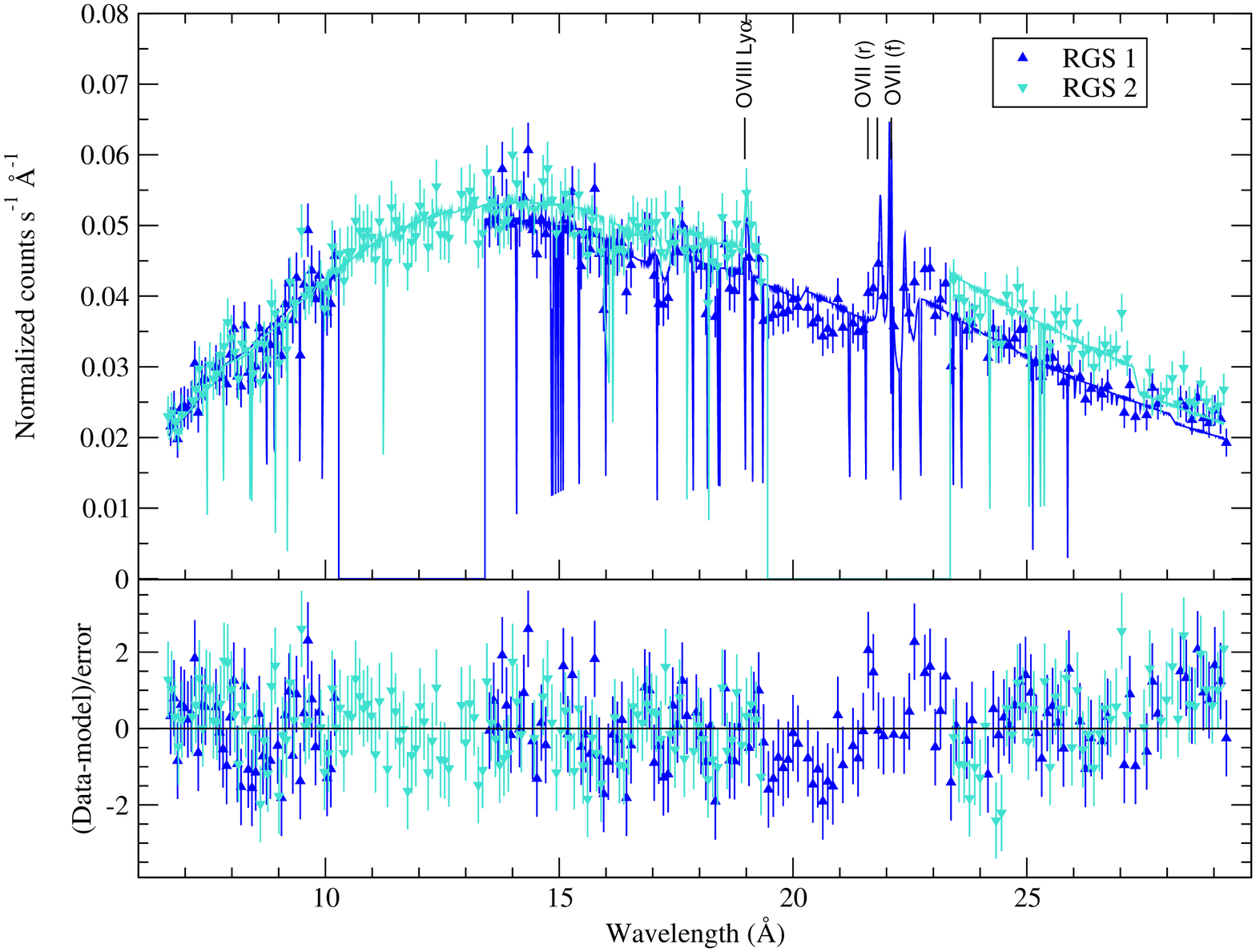}
\caption{As in Fig.\ \ref{eso:all-final} for \mkn.}
 \label{mkn:all-final}
\end{figure}

The model that best reproduces the EPIC spectra of \mkn leaves residuals on
the RSG spectra that could come from the presence of narrow emission lines. 
Also departures $2\sigma$ larger than zero are found at $\sim6.4$ keV that
point to the presence of a weak Fe line. Although the addition of this line
was found to not be statistically significant to the fit of the EPIC data, we
decided to incorporate it in the five-spectra fit. 
This line was modeled with a Gaussian profile whose width was left free.
As a result, a very broad feature ($\sigma\sim2$\,\kev) centered on
$\sim5$\,\kev was obtained. 
Since the feature seen in the residuals is not very wide, we tried to find 
the best parameters for this line ``by hand''. 
Then we realized that the line dispersion could be frozen to the instrumental
resolution. 
With the initial parameters 
found in this way we get a satisfactory fitting for this Fe
emission line at about 6.4\,\kev, although the fitting process is not able to
find the errors associated with the energy value of the line.
The narrow features that appear in the RGS data were modeled with a second 
O{\sc vii} triplet (i.e., another set of three Gaussian lines linked in
energies and dispersions) and an O{\sc viii} line. 
It should be emphazised that these lines were 
included only in the RGS model and that all of them turn out to be significant
to the fit. 
Figure \ref{mkn:all-final} shows the final model adopted
for \mkn, along with the corresponding errors.

The parameter values of the best-fitting model for each of these objects are
listed in Tables \ref{4obj:all-continuum} and \ref{4obj:all-lines}. 

\section{Discussion \label{sec:discussion}}

\subsection*{UV emission}
From the OM data of \he we have found that the IUE average spectrum obtained in
1990 is representative of its flux state during the {\em XMM-Newton}
observation. 
Using the {\em IUE} flux at 2500\,\AA\ and the flux at 2\,keV from
\pn\ data, $F(2\,keV)=7.67\times 10^{-12}$ \funitsk, we obtain an
optical/X-ray spectral index $\alpha_{ox}=1.2$ for this object. This value
is typical of Seyfert\,1 AGNs, as shown by \cite{1987ApJ...323..243Wilkes}
and \cite{2009MNRAS.392.1124Vasudevan}. 

In the case of \mkn, we have found that the {\em IUE} fluxes corresponding to
the effective wavelengths of the OM filters UVM2 and UVW1 ($\lambda$ 2310 and
2910\,\AA, respectively) seem to be lower than OM fluxes by 
$1\times10^{-14}$ \funitsa. Taking this correction into account we
estimated the flux at 2500\,\AA\ that should correspond to the 
{\em XMM-Newton} observation. Using this estimate and the flux at 2\,keV from
\pn\ data, $F(2\,keV)=7.28\times 10^{-12}$ \funitsk, we obtain an
optical/X-ray spectral index $\alpha_{ox}=1.2$ for this object, in very good
agreement with the one obtained by \cite{2009MNRAS.392.1124Vasudevan} using the
same X-ray data. 

\subsection*{Variability}
The light curves of all the four objects analyzed in this work show low
amplitude flux variations in the X-ray bands studied. 
\eso and \cts are the sources with the lowest count rates. 
\eso shows only marginally significant variability in the soft band and no
statistically significant variations in the hard band. The
soft and hard band light curves are statistically well represented by constant
fluxes with small variations of about 2 an 1 $\sigma$, respectively.  

\cts shows X-rays flux variations in both soft and hard bands, which are more
significant for the former ($3\sigma$).
Both light curves show a rather constant mean flux value.
Under a careful visual inspection of the shape of the light curves, there
seems to be an anti-correlation between the flux variations in the soft and
hard bands. This anti-correlation, if real,
could be interpreted in different ways \citep{1996ApJ...469L..37Ford}: \\ 
  (i) Thermal models \citep[e.g.,][]{1980A&A....86..121Sunyaev}: when the soft
X-ray flux increases, the Comptonizing plasma cools down, resulting in a
decreased temperature and hard X-ray flux. \\
  (ii) Non-thermal models: the soft and hard X-ray anti-correlation is the
result of the suppression of particle acceleration by inverse
Compton cooling caused by an enhanced soft X-ray background.

In \he and \mkn, flux variations in the hard band follow the corresponding ones
in the soft band with no lag between features in \he and a delay of about
1000\,s ($\sim$17 minutes) in the case of \mkn with the soft band variations
preceding those in the hard band. 
This result is in very good agreement with the work of   
\cite{2006ApJ...651L..13Dasgupta}, who did a detailed analysis of the
variability during the same {\em XMM-Newton} observation. 
They reported the discovery of X-ray delays 
ranging from a few minutes to an hour between the light curves in seven
selected spectral intervals (0.2-0.3, 0.3-0.42, 0.42-0.58, 0.58-0.8, 0.8-1.2,
1.2-2, and 2-12\,\kev). These authors explained the time lags as
the effect of inverse Compton scattering by highly energetic electrons (a hot
plasma; $T\sim 10^8-10^9$\,K, \citealt{1980A&A....86..121Sunyaev}) within
about ten Schwarzschild radii of the black hole.   

\subsection*{X-ray continuum emission}

Regarding the shape of the continuum emission of the objects studied here, a
power law accounts for the hard X-ray emission in all cases. 
The photon indices of \eso and \cts have standard values between 1.5 and 2.5
for Seyfert\,1s
\citep{1994MNRAS.268..405Nandra,2000MNRAS.316..234Reeves,2005A&A...432...15Piconcelli}.

Analyzing \textit{ROSAT} data of \eso (taken in 1990 with the source in a high
flux state, in the 0.1-2.4 keV range), 
\cite{2001A&A...367..470Grupe,2004AJ....127..156Grupe} fitted the soft X-ray
continuum with a power law with a photon index of $\Gamma\sim2.4$ that is
consistent with the presence of the soft excess component. 
In \cite{2010ApJS..187...64Grupe}, these authors analyzed \textit{SWIFT} data  
(taken in 2008 with the source in low state) reporting that
the continuum in the 0.3-10 keV range could be described with
a power law with a photon index of $\Gamma\sim1.7$. This value is consistent
with the hard power law index, i.e. with no soft excess. 
The \textit{XMM-Newton} data were also taken with the source in a low flux
state. However, in contrast to the findings of Grupe and collaborators, we do
not find the hard power law model sufficient for characterizing the continuum 
in the 0.3-10\,keV range. In our case, a contribution by a soft excess
component is found at energies lower than about 1\,keV.

\he and \mkn have slightly lower photon indices than the ones typical of
Seyfert 1s but they are still within the observed range of values for this
kind of objects
\citep{1994MNRAS.268..405Nandra,1997MNRAS.286..513Reynolds}. 
\cite{2009ApJ...690.1322Winter} fitted the 0.5-10\,\kev energy range of the
{\em SWIFT} XRT data of \he and the {\em ASCA} data of 
\mkn using single power laws with photon
indices of $\Gamma=1.92\pm0.1$ and $1.78\pm0.02$, respectively. 
In both cases the values derived are between the hard and soft power
law photon indices obtained by us (see Table \ref{4obj:all-continuum}).  
However, in our observations we do not find that single power law models can
be used to characterize the whole spectral range of the {\em XMM-Newton} data
of these objects. 

\subsection*{The Fe lines}
In these four objects, we can see that a power law alone does not 
satisfactorily model the shape of the spectra in the 2-10\,\kev range. 
In all of them, an emission feature arises around 6.4\,\kev. 
This is the characteristic energy of the fluorescent neutral iron line, which
is usually observed in the spectra of this kind of objects.
The iron emission line energy increases with increasing ionization. It is
6.4\,\kev for neutral Fe, 6.7\,\kev in the H-like Fe, and 6.966\,\kev in
the He-like Fe \citep{1999agnc.book.....Krolik}.
We used a Gaussian profile to characterize the energy, strength, and dispersion
of the observed feature. 
For all the objects we find that this line is statistically significant
to the fit, and the fitted energies are compatible -within 
the errors- with their being the neutral Fe\,K$\alpha$ line.

For the iron line dispersion, we find values of 
$0.07_{-0.04}^{+0.05}$\,\kev ($3300_{-1900}^{+2300}$\,\kms) for \he, 
$0.06_{-0.04}^{+0.04}$\,\kev ($2800_{-1900}^{+1900}$\,\kms) for \cts, 
and narrower than the instrumental resolution for \eso and \mkn.  
The equivalent widths (EW) of these lines are 
$0.16_{-0.12}^{+0.12}$\,\kev for \eso, $0.07_{-0.04}^{+0.04}$\,\kev for \he, 
$0.14_{-0.06}^{+0.06}$\,\kev for \cts, 
and $0.05_{-0.02}^{+0.02}$\,\kev for \mkn.

The {\em XMM-Newton} data of \he and \mkn were included in the
sample analyzed by \cite{2007MNRAS.382..194Nandra} to study the 
Fe\,K$\alpha$ line observed in 26 Seyfert galaxies. 
For the iron line identified in the \he data, they reported a central energy of 
$5.53_{-0.22}^{+0.14}$\,\kev (errors are given at the 68\% confidence
level). This energy is redshifted with respect to the 
laboratory energy of the neutral iron K$\alpha$ line.
Moreover, the line they found is broader 
($\sigma=0.14_{-0.14}^{+0.24}$\,\kev, i.e., about 7600\,\kms) than our
estimation, and the EW (of $30_{-21}^{+28}$\,eV) is about half our
value. All these reasons, and a careful inspection 
of Figure 3 of \cite{2007MNRAS.382..194Nandra} lead us to think that the
feature they fitted with a Gaussian profile is not the Fe\,K$\alpha$ line. 

In the case of \mkn, \cite{2007MNRAS.382..194Nandra} found the iron line at a
central energy of $6.53_{-0.11}^{+0.14}$\,\kev. Taking the errors into account,
this feature is compatible with the neutral fluorescent iron line at 6.4\,\kev,
hence also with our estimation. 
The width ($\sigma=0.00_{-0.00}^{+0.23}$\,\kev) of the 
line reported by these authors is in good agreement with our value, and their
EW ($17_{-12}^{+41}$\,eV) is about 2.5 times lower than our estimation, but
is also compatible within the errors. 
\cite{2007A&A...465...87Boller} have also analyzed the same {\em XMM-Newton}
observation of \mkn. They found similar
parameters for the Fe line profile: a narrow line (with the line width
unresolved within the energy resolution of the \pn detector), a central energy
of $6.4_{-0.04}^{+0.02}$\,\kev, and an EW of $52_{-16}^{+18}$\,eV.
Our line parameters are also in good agreement with those more recently given 
by \cite{2009A&A...507..159DeMarco}: E$_{\mathrm{Fe}}=6.41\pm0.04$\,\kev, 
$\sigma<0.13$\,\kev, and EW$=46\pm13 1.4$\,eV.

In the \eso spectra, we have also identified a second narrow line in the hard
band with an energy of $\sim6.97$\kev, probably  
originated by highly ionized iron. This line has not been previously identified
in the spectra of this source, but it has been reported in other Seyfert 1s,
e.g., by \cite{2007A&A...470...73Longinotti} in MKN\,590 and \cite{2008MNRAS.389L..52Bianchi} in NGC\,7213.

\subsection*{Soft excess}

The extrapolation of the hard X-ray models of the four sample objects to
lower energies, down to 0.35~keV, reveals the existence -in all of them- of a
soft-X-ray flux in excess of this extrapolation, from $\sim\,2$\,\kev and
below.  
This excess was previously reported only for \mkn 
\citep{2007A&A...465...87Boller}.
After testing the most typical components used to model the soft excess, we
find that in \eso and \cts it can be accounted for with a black body component
with $kT=0.08_{-0.04}^{+0.03}$ and $kT=0.126_{-0.004}^{+0.004}$\,keV,
respectively. On the other hand, for \he and \mkn, a 
(soft) power law is the model that best accounts for the observed soft excess
with photon indices of $2.62_{-0.03}^{+0.03}$ and $2.59_{-0.09}^{+0.34}$,
respectively. 
Both the values of the $kT$ of the black bodies and the slopes of the power
laws found by us are consistent with the values found in many other Seyfert
galaxies \citep{2005A&A...432...15Piconcelli}. 

\cite{2001A&A...367..470Grupe} analyzed the {\em ROSAT} data
of \eso in the 0.2-2\,\kev energy range. These authors use a power law with a
photon index $\Gamma=2.41\pm0.07$ to describe the continuum
emission in the soft band. This is a characteristic value for AGNs with a
small-amplitude soft excess as is the case of \eso. In a recent
work, \cite{2010ApJS..187...64Grupe} using two {\em SWIFT} observations data
in the same soft band (0.2-2\,\kev) found soft power law photon indices of
$1.72\pm0.10$ and $1.60\pm0.14$, which are more
typical of the spectral indices observed in the 2-10\,\kev energy band. 
This behavior could indicate that the soft excess is variable both
in flux and shape as seen in NLS1 objects.

{\em ROSAT} data of \mkn were also analyzed
by \cite{2001A&A...367..470Grupe}. They found a photon index of $2.29\pm0.09$
for the energy range of 0.2-2\,\kev, slightly lower than our
value. Using a more complex model than ours to analyze the same observation,
\cite{2007A&A...465...87Boller} also found a good description of the soft
X-ray excess with a power law 
with a photon index of $2.5_{-0.02}^{+0.13}$, which is in very good agreement
with the value we find.
It must be noticed that another component of their model, mainly a
starburst, also contributes to the soft X-ray emission, as discussed below.

\subsection*{Broad soft X-ray features}

After the EPIC spectra of \he and \mkn were modeled with two power laws and an
Fe Gaussian component at about 6.4\,\kev, the residuals of the fit
clearly show that the models underestimate the counts around 0.5-0.6\,\kev.
The residuals have a Gaussian-like shape and arise in the range in
which the O{\sc vii} laboratory energies lay (0.5740, 0,5686,
0.5610\,\kev). These 
energies are also covered by the \rgs spectrometers, so the features seen
in the EPIC soft X-rays range should also be seen in the RGS spectra. 

For \he, a Gaussian profile ($\sigma=0.056_{-0.009}^{+0.010}$\,\kev,
i.e., $29000_{-5000}^{+5000}$\,\kms)
accounts for the broad feature seen in the EPIC 
and \rgs spectra with a central energy of $0.578_{-0.007}^{+0.006}$\,\kev 
($21.5_{-0.2}^{+0.2}$\,\AA). This line energy is compatible within the errors
with the resonant line of the O{\sc vii}. The \rgs data does not show any
other significant feature that indicates emission or absorption.

In the case of \mkn, a similar single broad Gaussian profile does not account
for the observed feature either in the EPIC data or in the \rgs data. 
To achieve a good description of the broad and narrow
emission features that arise in the EPIC and \rgs spectra, respectively, we
have included in the model some Gaussian profiles.  
Three broad profiles account for the broad excess in the EPIC data. Since we
assume that these lines are produced by the O{\sc vii}-He$\alpha$ triplet, we
fixed their relative energies to the laboratory values. In this way we find the
lines at energies of $0.59_{-0.16}^{+0.04}$, $0.5872$, and $0.5796$\,\kev 
(21, 21.1 and 21.4\,\AA).
The line widths were also forced to be the same for the three lines obtaining 
dispersions of $\sigma=0.04_{-0.02}^{+0.02}$\,\kev, equivalent to
$20000_{-12000}^{+10000}$\,\kms. 
Regarding the width and intensities, we can see that the lines 
overlap and the third one dominates the profile, although we
notice that the normalizations are unconstrained (Table \ref{4obj:all-lines}).
Line energies are highly blueshifted, $z\sim-0.034$, when taking the \mkn
redshift into account ($z\sim0.035$). Since these two values are almost equal
(in absolute value), this implies that the broad feature, if it is associated
with O{\sc vii}, is at rest relative to the observer (before the redshift
correction was performed), and it could be due to calibration uncertainties.

\cite{2007A&A...465...87Boller} modeled the broad EPIC feature with only one
Gaussian profile and a collisionally ionized plasma. For the line, they found 
a central energy of $0.554_{-0.003}^{+0.002}$\,\kev and a width of
$0.013_{-0.003}^{+0.004}$\,\kev. They concluded that this line is originated
in the broad line region. 
Even when the central energy found by Boller
and collaborators is compatible -considering the errors- with our
estimation, the line they found is narrower than our lines. 

\subsection*{High-resolution spectral features}

We used narrow lines (widths fixed to the instrumental
resolution) to fit the narrow emission
features seen in the \rgs spectra of \mkn. 
To do this we introduced a second triplet and a single Gaussian to model the
lines that are probably related to emission from O{\sc vii} and O{\sc viii},
respectively. Again, we fixed the relative energies of the triplet lines to
the laboratory values. For the O{\sc vii} triplet we find energies of about
$0.5669$, $0.5615$, and $0.5539$\,\kev (21.9, 22.1, and 22.4\,\AA;
unconstrained values), 
implying a velocity of about $3700$\,\kms ($z\sim0.012$). 
The flux errors in the O{\sc vii} lines do not allow us to
obtain the coefficients that are commonly used to estimate the density and
temperature of the emitting gas
\citep{1969MNRAS.145..241Gabriel,2000A&AS..143..495Porquet}. 
The O{\sc viii} line is located at a central energy of
$0.652_{-0.005}^{+0.004}$\,\kev ($\sim19$\,\AA), which is compatible with being at
the rest frame of \mkn.
We do not find any other line significant to the fit. 

\cite{2007A&A...465...87Boller} also used narrow lines to model the narrow
features. They model the same features associated with the oxygen plus
two more lines from N{\sc vii}-Ly$\alpha$ 
(E$_{\mathit{lab}}=0.4994$\,\kev, i.e., $\lambda_{\mathit{lab}}=24.8$\,\AA)
and C{\sc vi} Ly$\alpha$ 
(E$_{\mathit{lab}}=0.3668$\,\kev, i.e., $\lambda_{\mathit{lab}}=33.8$\,\AA). 
The line parameters given by them for the oxygen lines are
compatible -within the errors- with our estimations. We find neither the 
N{\sc vii} nor C{\sc vi} lines.

\eso \rgs data show a few emission signatures that deviate from the
comprehensive five-spectra continuum fit. However, they are not 
statistically significant when modeled with Gaussian functions. 

\cts is studied in detail in the X-ray band here by the first time. 
We find one statistically significant emission line in the
soft band when the comprehensive five-spectra fit is performed. This line can
be identified as produced by O{\sc vii}. The line found has a central energy of
$0.574_{-0.006}^{+0.003}$\,\kev ($\sim21.6$\,\AA), which is very similar to
the energy of the recombination component of the O{\sc vii} triplet and has a
dispersion of about 0.005\,\kev ($\sim2600$\,\kms).

Only in the case of \mkn and \cts do we see a feature between 15 and 17\AA\ in
absorption that can be associated with the Fe-L UTA (Unresolve Transition
Array), which is the characteristic signature of warm absorbers
\citep{2001A&A...365L.168Sako}.
We tested for the presence of warm absorption related with these sources using
PHASE \citep{2003ApJ...597..832Krongold}, but in neither case
have we found a satisfactory result that would inprove the model fitting.


\section{Summary and conclusions  \label{conclusions}}

We analyzed all the information provided by the {\em XMM-Newton}
satellite, including the UV range, of the four Seyfert 1 galaxies: \eso, \he,
\cts, and \mkn. 
Our main effort, however, focused on the comprehensive analysis of
the X-ray data that allows the characterization of the observed features in
each object. 

All the galaxies were found to vary on long time scales of
a few years, i.e. when comparing our observations with those from other
missions.
Variations by factors as high as 10 or more in the soft X-ray fluxes of two of
the galaxies 
(\eso and \cts) were detected, with the {\em XMM-Newton} observations occurring in
the low state. 
For \he and \mkn, factors of only about 2.5 and 2, respectively, are measured
for the amplitudes of the variations. 
This probably means that they were still in a high-luminosity state
when observed by {\em XMM-Newton}, so they are significantly brighter
than the other two (by about an order of magnitude). The detected amplitude
of the variations in the hard X-rays are lower or none, sometimes owing to
lack of data in the highest soft X-ray state.
In the UV, we compared the OM fluxes with previous UV values from the {\em
IUE} satellite, when available. The variability amplitude is low, with
factors of about 2. 

On short time scales, we detected correlated soft and hard X-ray variability
in \he and \mkn, those galaxies with the highest signal-to-noise ratio data,
with \mkn showing a 1000\,s delay between the soft and hard X-rays.

The continuum of the four AGN studied can be fully characterized by a single
power law in the hard X-ray range, i.e. from 2 to 10\,keV, as in most other
Seyfert\,1 galaxies and type 1 QSO. For all four galaxies, an extra component
is needed in the soft X-rays, often called the soft-X-ray excess, and found as
well in the great majority of QSOs. This soft excess is successfully modeled
with a black 
body in two of our galaxies (\eso and \cts) and with a steep power law
in the other two (\he and \mkn). The parameters that define all the continuum
components have standard values, although in the two brightest objects the
index of the hard power law continuum lies at the lower end of its
distribution, with 
values $\Gamma = 1.2 - 1.4$. The soft excess black body temperatures,
$kT_{BB} \sim 80 - 130$\,eV, are well within the ranges found in other
Seyfert\,1 and QSO, and so are the spectral indices of the two soft power
laws.

The above standard models and parameters seem to indicate that the
standard explanation for the continuous X-ray emission is valid for these
galaxies: 
the soft photons originating in the accretion disk are Compton up-scattered by
hot electrons with a thermal distribution, probably located in a corona above
the accretion disk that would be responsible for the power-law hard X-ray
component of the spectrum. The origin of the soft X-ray emission, as 
already argued by several authors, is more difficult to understand, since the
maximum temperature from an accretion disk is expected to be below the above
values. 

No intrinsic absorption by cold material is required to explain
the observations in either of our galaxies. Interestingly,
none of these objects shows clear signs of partially ionized (warm)
material in the vicinity of the central black hole and in the line of
sight. 
Considering what was found for UGC\,11763, the fifth object of our
study, which was analysed in a previous paper
\citep{2009A&A...505..541Cardaci}, 
it is interesting to note that the ionized
material was detected in this galaxy, which was one of the two weakest
objects. This means that the lack of detection in the other galaxies is due to
intrinsically weaker or absent absorption in the line of sight and not to a
lower signal-to-noise ratio in the data.
This is apparently at odds with the expectation that about half of the
Seyfert\,1 galaxies should show evidence of this warm absorption. 
Nonetheless, given the low number of galaxies
studied, our results are fully compatible with this expectation.
This is even truer if we consider that two more spectra (\cts and \mkn)
show hints of weak absorption, although not significant ones. 

With the exception of the well-known Fe-K$\alpha$ line at 6.4\,keV, detected
in all the four objects, not many emission features have been detected in
spite of our detailed analysis of RGS data 
of a reasonably quality. The detected iron K$\alpha$ lines are
generally weak, sometimes even only marginally detected, and with equivalent
widths as low as 50 and 70\,eV in the two brightest galaxies, \mkn and \he,
respectively. The Fe-K$\alpha$ lines are not significantly 
broader than the EPIC spectral resolution. Fe{\sc xxvi} at 6.97\,keV is
detected in one of the galaxies (\eso).
The O{\sc vii}-He$\alpha$ triplet in the soft 
X-rays is detected with instrumental RGS resolution in two of the four
galaxies (\cts and \mkn), although not all the triplet components are always
found. It is detected as a broad line in two of the galaxies, the two
brightest, and, finally the O{\sc viii}-Ly$\alpha$ emission is detected in
only one of the galaxies, \mkn. 

If they could be extrapolated to the Seyfert\,1 galaxies as
a class, the above results would mean that only a few and very weak emission
features can be 
expected in their X-ray spectra.
This is in line with what has been reported in the literature.
That we detected several of our galaxies in a low state means that the absence
of clear emission lines cannot be fully attributed to dilution of those lines
by a strong continuum. 


\medskip

\paragraph{Acknowledgments}
This research is completely based on observations obtained with 
{\em XMM-Newton}, an
ESA science mission with instruments and contributions directly funded by ESA
Member States and NASA. For the spectral fitting, software
provided by the Chandra X-ray Centre (CXC) in the application package Sherpa
was used. 

This work has been supported by DGICYT grant AYA2007-67965-C03-03. MC
acknowledge support from the Spanish MEC through FPU grant 
AP2004-0977. Furthermore, partial support from the Comunidad
de Madrid under grants S-0505/ESP/000237 (ASTROCAM) and 
S-0505/ESP-0361 (ASTRID) is acknowledged.
MC and GH also acknowledge the hospitality of the UNAM.
YK acknowledges support from the Faculty of the European Space Astronomy
Centre (ESAC) and the hospitality of ESAC.

\bibliographystyle{aa}
\bibliography{16198}
\end{document}